\documentclass[a4paper,fleqn,longmktitle]{cas-dc}

\pdfoutput=1

\usepackage[authoryear]{natbib}

\usepackage[british]{babel}
\usepackage[english=usenglishmax]{hyphsubst}
\usepackage{microtype}
\newcommand{\earth}{$_\oplus$}
\newcommand{\sol}{$_\odot$}
\def\rfnce{\par\noindent\hangindent 20pt {}}

\begin{document}
\let\WriteBookmarks\relax
\def\floatpagepagefraction{1}
\def\textpagefraction{.001}
\shorttitle{Models of the {\it in situ} formation of detected extrasolar giant planets}
\shortauthors{P.~Bodenheimer, O~Hubickyj, \& J.J.~Lissauer}

\title [mode = title]{Models of the {\it in situ} formation of detected extrasolar giant planets}\tnotemark[1,2]

\tnotetext[1]{UCO/Lick Observatory Bulletin No. 1389.}
\tnotetext[2]{Paper presented at Protostars and Planets IV Conference, Santa Barbara,CA, July 1998.}

\author[1]{Peter Bodenheimer}[orcid=0000-0001-6093-3097]
\ead{peter@ucolick.org}
\address[1]{UCO/Lick Observatory,
Board of Studies in Astronomy and Astrophysics, University of California, Santa Cruz, CA 95064, USA}

\author[1,2]{Olenka Hubickyj}[]
%\ead{}
\address[2]{Space Science Division, NASA-Ames Research Center, Moffett Field, CA 94035, USA}

\author[2]{Jack J.\ Lissauer}[orcid=0000-0001-6513-1659]
%\ead{Jack.J.Lissauer@nasa.gov}
%\address[2]{}

%\cortext[1]{Corresponding author}

\begin{abstract}
We present numerical simulations of the formation of the planetary 
companions to 47 UMa, $\rho$ CrB, and 51 Peg. They are assumed to have formed
{\it in situ} 
according to the basic model that a core formed first
by accretion of solid particles, then later it captured substantial amounts
of gas from the protoplanetary disk. 
In most of the calculations we prescribe a constant accretion rate 
for the  solid core. 
The evolution of the gaseous envelope
is calculated according to the following assumptions: (1) it is in 
quasi-hydrostatic equilibrium, (2) the gas accretion rate is determined
by the requirement that the outer radius of the planet is the place at 
which the thermal velocity of the gas allows it to reach the boundary
of the planet's 
Hill sphere, (3) the gas accretion rate
is limited, moreover, by the prescribed maximum rate at which the nebula 
can supply the gas, and (4) the growth of the planet stops 
once it obtains approximately the minimum mass determined from radial
velocity measurements (in one case the planet is allowed to grow to
twice this limit).  Calculations are carried out
through an initial phase during which solid accretion dominates, past
the point of crossover when the masses of solid and gaseous material
are equal, through the phase of rapid gas accretion, and into the
final phase of contraction and cooling at constant mass. 
Alternative calculations are presented for the case of 47 UMa in which
the solid accretion rate is calculated, not assumed, and the 
dissolution of planetesimals within the gaseous envelope is considered.
In all cases there is a short phase of high luminosity (10$^{-3} - 10^{-2}$
L$_\odot$) associated with rapid gas accretion.  The height and duration of
this peak depend on uncertain model parameters. The conclusion is
reached that {\it in situ} formation of all of 
these companions is possible
under some circumstances. However, it is more likely that orbital
migration was an important component of the evolution, at least for 
the planets around $\rho$ CrB and 51 Peg. 
\end{abstract}

\begin{keywords}
origin of planetary systems\sep
 jovian planets, formation\sep
jovian planets, interiors\sep
accretion\sep
extrasolar planets

\vspace*{10mm}
Received:~October 16, 1998\sep
Revised:~April 16, 1999\sep

\vspace*{5mm}
Citation:~Icarus, Volume 143, Issue 1, January 2000, Pages 2-14\sep
\url{https://doi.org/10.1006/icar.1999.6246}

\end{keywords}

%\begin{graphicalabstract}
%\resizebox{0.5\linewidth}{!}{%
%\includegraphics{grabs1}}\\
%\resizebox{0.5\linewidth}{!}{%
%\includegraphics{grabs2}}\\
%\resizebox{0.5\linewidth}{!}{%
%\includegraphics{grabs3}}
%\end{graphicalabstract}

%3 to 5 bullet points (maximum 85 characters, including spaces, per bullet point)
%\begin{highlights}
%\item Jupiter's formation is modeled via core-nucleated accretion
%\item Solids' accretion accounts for interactions with an evolving disk of planetesimals
%\item Evolution within a dispersing nebula is approximated via simplified disk models
%\item Structure calculations are performed throughout the planet's history, for $5$~Gyr
%\item Final radii and luminosities agree with those of Jupiter within $10$\%
%\end{highlights}

\maketitle

\microtypesetup{activate=true}

\section{INTRODUCTION}

The recent discoveries of giant planets orbiting at relatively close
distances around solar-type stars (reviewed by Marcy 
and Butler 1998) have led to intense speculation 
regarding their origin. The objects discovered so far have orbits
significantly closer to their star than Jupiter's distance
from the Sun, but have masses in the range 0.4 -- 10 Jupiter masses (M$_J$).
Examples of objects with nearly circular orbits, but with 
distances from the
central object great enough  so that 
the circularity of the orbit is unlikely to have been
achieved through the influence of the
tidal effects of the star, include the companion to 47 UMa (Butler
and Marcy 1996) at 2.1 AU
and the companion to  $\rho$ CrB   (Noyes {\it et al.} 1997) 
at 0.23 AU.  Examples of objects 
very close to the central star, with periods of only
a few days, include  the companions to 51 Peg (Mayor and Queloz 1995),
55 $\rho^1$ Cancri (Butler {\it et al.} 1997), $\tau$ Boo 
(Butler {\it et al.} 1997), 
and $\upsilon$ And (Butler {\it et al.} 1997). 
A few of the newly-discovered objects travel on eccentric orbits.       
In the cases of HD 114762 (Latham {\it et al.} 1989) and 
70 Vir (Marcy and Butler 1996), the planets are relatively massive with  9 and
6.6 M$_J$ as minimum masses, respectively. In the case of the companion
to 16 Cyg B (Cochran {\it et al.} 1997),  the eccentricity is 0.6 but the minimum
mass is only 1.7 M$_J$.

A number of theories have been proposed regarding the origin of these 
systems. The objects could have formed by fragmentation of a collapsing 
protostellar cloud and subsequent orbital interactions among the 
fragments. 
The companion to Gl 229 
(Oppenheimer {\it et al.} 1995; Nakajima {\it et al.} 1995) 
almost certainly falls into this class, because of its relatively
high mass ($\approx 40$ M$_J$; see Allard {\it et al.} 1996; Marley {\it et al.} 1996), 
and Black (1997) argues that most of the objects discovered
in radial velocity surveys do also.
This mechanism is most likely to produce
masses $>7$ M$_J$ (Low and Lynden-Bell 1976; Rees 1976) although it has not 
been definitely proved that smaller masses are impossible. 
The second mechanism
involves collapse of a molecular cloud core into 
a protostar and a disk, followed by gravitational
instability in the disk leading to fragmentation on a dynamical
time scale (Kuiper 1951; Cameron 1978; DeCampli
and  Cameron 1979). 
Numerical calculations on gravitationally 
unstable disks  by Adams and Benz (1992) and Boss (1997)
suggest that if a disk can be produced with appropriate conditions, 
fragmentation into objects of $\sim$ 10 M$_J$ rather than $\sim$ 1 M$_J$ 
is likely to occur. Laughlin and Bodenheimer (1994) argue that the 
collapsing system will not result in a disk that is subject
to fragmentation, but rather in one that develops a spiral wave pattern
that saturates in amplitude.  This second mechanism also 
has difficulties because, although it is possible  
to get a gravitationally unstable 
disk model with a mass of about  0.14 M$_\odot$ out to 10 AU
(Boss 1998),  the implied total disk mass
is high compared with most observations. 
Nevertheless, the more massive objects are candidates for this process.

The third process proposed for the formation of giant planets  is gradual
accretion of small solid particles in a low-mass protoplanetary disk, followed
by gravitational capture of gas. The main physical effects are reviewed
by Lissauer (1993). 
The stages that are envisioned are as follows:
(1)  Accretion of dust particles (Safronov 1969)  results in a solid core with mass $M_Z$ 
of  several M$_\oplus$, accompanied by a  gaseous envelope of very low mass, $M_{XY}$.
(2) Further accretion of gas and solids results in the mass of the 
envelope increasing faster than that of the core until a 
crossover mass ($M_{cross} \equiv M_Z = M_{XY}$) is 
reached (Perri and Cameron 1974; Mizuno 1980).
(3) Runaway gas accretion occurs with relatively little
accretion of solids, and the peak luminosity reaches 
$10^{-3} - 10^{-4}$ L$_\odot$ (Bodenheimer and
Pollack 1986).
(4) Accretion is terminated by tidal truncation (Lin and Papaloizou
1979, 1985, 1993)  or dissipation
of the nebula.
(5)  The planet contracts and cools
at constant mass to the present state
(Bodenheimer and
Pollack 1986; Saumon {\it et al.} 1996; Burrows {\it et al.} 1997).
The resulting orbits are likely to be initially relatively circular
and the masses not more than a few  M$_J$, because tidal truncation
sets in at that point and stops 
appreciable further accretion of gas (Lin and Papaloizou 1979). Note that the 
value of this limiting mass depends on uncertain disk
parameters such as viscosity. Moreover,  Artymowicz (1998) and Artymowicz
and Lubow (1996) have made calculations which indicate that accretion onto
a protoplanet can continue even after gap formation at a rate 
which depends on the disk viscosity; thus, the issue
of the limiting mass of a protoplanet formed by this process is controversial. 

Evolutionary calculations based on this general scenario have been performed
by Bodenheimer and Pollack (1986) through stages (1)--(5), under the
assumption of a constant solid accretion rate, and by Pollack {\it et al.}
(1996) through stages (1)--(3) with more detailed physics, including a 
(non-constant) solid accretion rate calculated from three-body accretion 
cross sections. These studies show that the core accretion theory is successful
in the sense that it
(i)   gives a mass in the solid component of the giant planets which
agrees with the observations of the giant planets in the Solar System, and 
(ii)  shows that the mass of  the solid component is relatively independent of
the position of the planet in the solar nebula, again in agreement with
the observations of Jupiter and Saturn. As discussed by Pollack {\it et al.} 
(1996), the theory must allow for the partitioning of the solid
component into an actual core and as dissolved 
material into the gaseous envelope; recent observational constraints  are
discussed by Guillot {\it et al.} (1997) and Wuchterl {\it et al.} (1999).   

An often-discussed  problem with the core accretion  process is that
in a `minimum mass' solar nebula the accretion times for
the cores of Jupiter and Saturn are too long, over 10$^7$ yr.
The range of observed disk lifetimes, although very uncertain, is quoted
to be $\sim 0.1-10$ Myr (Strom {\it et al.} 1993). 
However, if the disk's  surface density is increased by
a factor of a few over the minimum value, formation times are reasonable
(Lissauer 1987; Pollack {\it et al.} 1996) and 
the corresponding disk masses are still in
agreement with those deduced from observations (Beckwith and Sargent 1993).
Also, the excess mass could be invoked to explain the Oort cloud. 
Another problem  with the core accretion mechanism has arisen as a result of
hydrodynamic calculations by Wuchterl (1991). He found that at the    
beginning of stage (3), when rapid gas accretion starts, most of
the gaseous envelope is ejected, leaving only about 1 M\earth~around
the solid core. If the nebular density is increased to the point
where the planetary gaseous envelope becomes predominantly convective,
then the instability is avoided (Wuchterl 1995);
however, the required density is about a factor 8 
higher than  densities  in a `minimum mass' nebula at 5~--~10 AU from the
central star.  On the other hand, Tajima and Nakagawa (1997) find that the 
protoplanetary envelopes are dynamically stable to small perturbations
even in a minimum-mass nebula.

Lin {\it et al.} (1996) suggested that the companion
to 51 Peg formed by this mechanism at a distance of a few AU from
the star  and then migrated to its present
position through interaction with the disk. At distances of $\approx 
0.05$ AU (but not for appreciably greater distances) the migration
can be stopped by tidal interaction with the star or truncation of
the inner disk by the stellar magnetic field. More detailed calculations
of post-formation orbital evolution (Trilling {\it et al.} 1998; Murray 
{\it et al.} 1998) 
suggest that several of
the orbital positions of extrasolar planets can be explained. 
Certainly 
migration is a potentially  important element in the formation of all types  
of  giant planets (Goldreich and Tremaine 1980; 
Lin and Papaloizou 1986a,b; Lin 1997; 
Takeuchi {\it et al.} 1996; Ward 1997a,b). In this context, 
the planets with eccentric orbits
could be explained by various processes of excitation of eccentricity: 
interaction with the disk (Artymowicz 1992), 
close encounters  and scatterings between
two or more giant planets (Lin and Ida 1997; Levison {\it et al.} 1998; 
Weidenschilling and Marzari 1996; see also Rasio and Ford 1996
for a related explanation of 51 Peg type  planets), or 
the presence of a  distant binary companion as   in the
case of 16 Cyg B b 
(Mazeh {\it et al.} 1997;  Holman {\it et al.}  1997). 
However, the relatively 
high masses  of 70 Vir b and HD 114762 b cause some difficulty for their
formation by the core accretion process 
unless the standard tidal truncation limit
can be exceeded, and they may possibly have 
formed by fragmentation.
 
The calculations reported here 
are based on the third   process just described,  that is,
the standard mechanism for forming giant planets in a disk.
However, we do not include  post-formation 
migration of the planets in the calculations, but ask 
instead whether {\it in situ} formation is a viable possibility and
then consider the consequences. 
As examples we consider three objects with nearly circular orbits: 
51 Peg b, 47 UMa b, and $\rho$ CrB b. 
There are several good reasons why it is difficult to form a
giant   planet close to a star.  First, according to many nebular models
(e.g., Bell {\it et al.} 1997),
the temperature is too high at $\approx 0.05$ AU
from the central star to allow for condensation of solid
particles. Second, even if solid particles could condense, there is
insufficient mass in the inner region of the nebula to produce
a Jupiter mass.  Third, even if a Jupiter mass were to form, tidal
interactions between it and the disk would rapidly cause it to migrate
into the star (Goldreich and Tremaine 1980; Ward 1997a,b). 
The first objection does not apply to the 
system of $\rho$ CrB at 0.23 AU; for 51 Peg we use disk models which
have relatively low mass accretion rates and therefore are         
cool enough at 0.05 AU for
some condensible material to exist (Bell {\it et al.} 1997). The second
objection is overcome under the assumption that 
solid particles 
can migrate through the disk as a result of its normal (viscous) evolution
or gas drag, 
and that larger chunks can migrate relative to the disk as a result of
tidal effects.  Thus, for 
51 Peg b and $\rho$ CrB b we actually consider migration in the 
sense that we assume that solid material is delivered at a constant
rate to the formation site.
The third objection is still a major problem for all giant planets
and we do not consider it here, except that
in the case of 51 Peg b we consider an   alternative
scenario of {\it in situ} formation based on a modified disk structure
with reduced density in the inner region. 
This alternative is a modification of and 
elaboration upon a model for 51 Peg b recently presented by Ward (1997a),  
which involves the buildup of an inner planet (at 0.05 AU) through collisions 
of protoplanetary 
cores  in the range 1~--~10 M\earth~which have migrated there from the
outer parts of the disk. 
In the case of 47 UMa b (at 2.1 AU), the assumption of {\it in situ}
formation is certainly a reasonable one, and we consider two 
possibilities, first,  a model with variable solid accretion rate (as in 
Pollack {\it et al.} 1996), and second, one with constant solid 
accretion rate (as in Bodenheimer and Pollack  1986).
Other {\it in situ} models for the formation of 51 Peg b have 
been presented by Wuchterl (1997) and Ruzmaikina (1998); also, 
preliminary results of the present authors are found in 
Bodenheimer (1998).

\section{PHYSICAL ASSUMPTIONS AND METHOD}

\begin{table*}[!ht]
\centering
\resizebox{0.8\linewidth}{!}{%
\includegraphics[trim={2cm 14cm 2cm 2cm},clip]{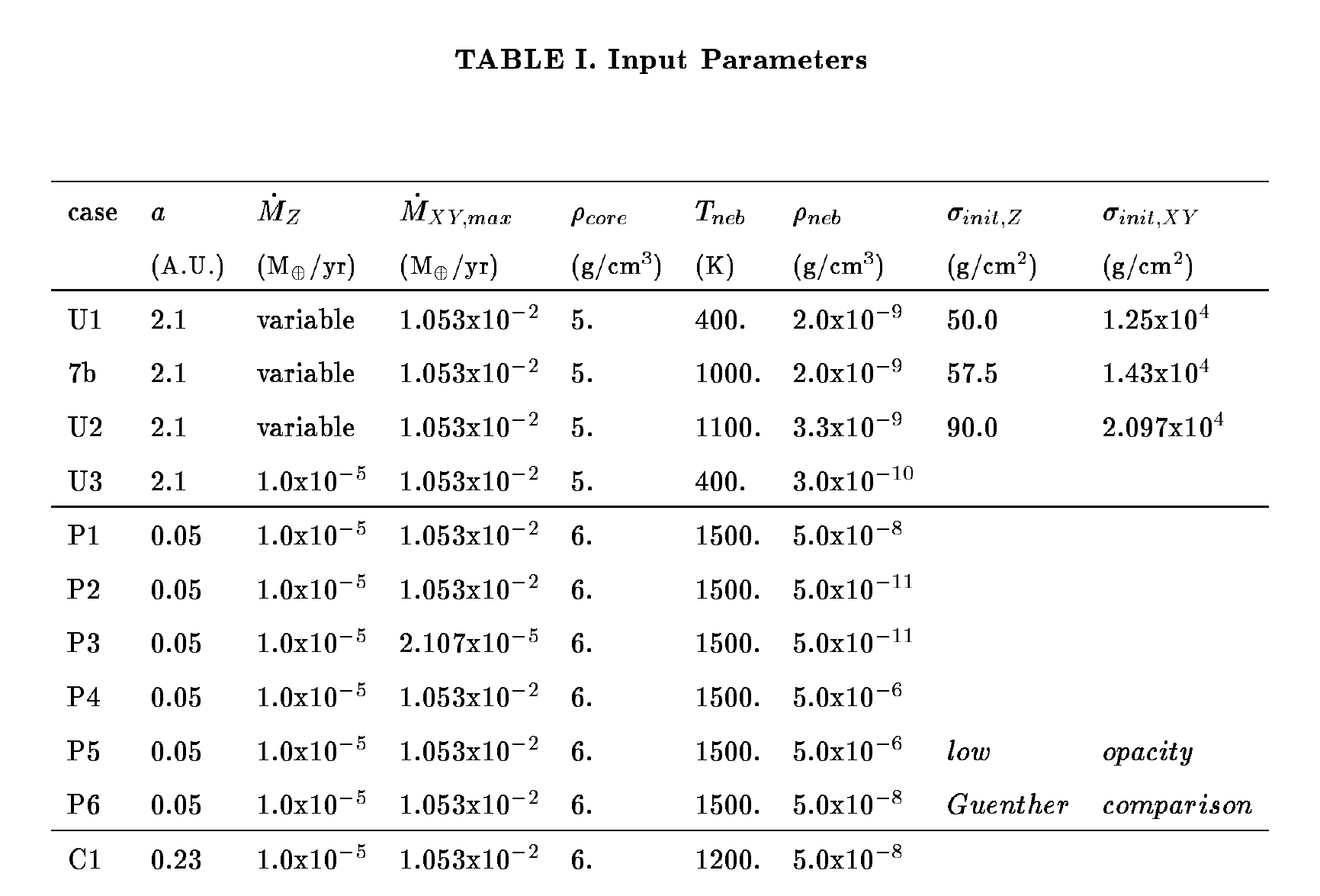}}
\end{table*}

In our model, the protoplanet consists of a solid core and a 
gaseous envelope, both of which evolve as a consequence of accretion. 
The computational procedure is described in detail in Pollack 
{\it et al.} (1996)
and consists of three main components: (1) the calculation of the rate
of accretion of solid material, assumed to be in the form of 
planetesimals,  onto the protoplanet, taking into account
the physical cross section as well as the gravitational enhancement
factor; (2)  the calculation of trajectories and destruction rate 
of planetesimals as they
pass through the gaseous envelope, to determine the radial profile of
energy deposition in cases where the planetesimals do not plunge all
the way to the core; and (3) a calculation of the evolution and 
mass accretion rate of the gaseous envelope, under the assumption that
the planet is spherical and that the standard equations of stellar structure
apply. Table I lists the parameters for each of the cases 
computed.  A glossary of symbols is provided in the Appendix. 
In Cases U1 and U2   the full procedure is used. In all other 
cases a simplified procedure is used
in which the solid accretion is assumed to occur at a constant, 
parameterized, rate and in which the accreted solids are 
assumed to fall directly to the core, releasing all of their
gravitational energy there. 

For the calculation of the solid accretion rate (Cases U1, U2) 
the basic assumptions
are: the protoplanet is surrounded by a disk with an initial  uniform surface
density $\sigma_{init,Z}$ of solid material, and the solid material is in the
form of planetesimals of uniform size, 100 km in radius. The rate of 
growth of  the solid core of a protoplanet is given by the standard
expression (Safronov 1969) 

$$ { \dot M_Z } =  {\pi R_c^2 \sigma_Z \Omega F_g} ~, \eqno(1) $$ 
where $\Omega$ is the orbital frequency, $R_c$ is the effective 
capture radius of the protoplanet, and $F_g$ is the gravitational
enhancement factor. Values for $F_g$  are obtained from the simulations
of Greenzweig and Lissauer (1992). The feeding zone of the planet is
taken to extend to about 4 Hill-sphere radii away from the protoplanet
in each direction (Kary and Lissauer 1994). Within the current feeding
zone, the surface density, $\sigma_Z$, is reduced appropriately as material
accretes onto the protoplanet. The feeding zone grows as a result of
the increase of the planet's mass; migration of external planetesimals into 
the feeding zone is not considered.

The second component of the calculation (Cases U1, U2) 
involves the interaction of 
planetesimals with the envelope, through which the energy
deposition as a function of envelope mass fraction and the effective
capture radius, $R_c$, are determined. If the envelope mass is small,
planetesimals simply pass through almost unaffected and land on the
core, releasing all of their kinetic energy there. But if the envelope
mass is more than a few M\earth, gas drag, ablation, and vaporization 
affect the trajectory and can result in mass and energy deposition in
layers above the core. The calculation (Podolak {\it et al.} 1988) 
of planetesimal trajectories through the envelope takes these  effects 
into account, and the energy deposition profile is supplied to
the program that solves the envelope structure equations (below). 
The mass is assumed to be added to the core. 
The value of $R_c$ is derived from the critical impact parameter, 
determined from the trajectories for which
the encounter results in the escape 
of a planetesimal rather than a capture. 
The details of the calculation of planetesimal trajectories
are provided by Pollack {\it et al.} (1996). 

The third component involves the calculation of the structure and
evolution of the envelope; it is included in all cases. 
The structure is determined from the equations of 
mass conservation, energy conservation, hydrostatic equilibrium, 
and radiative or convective energy transport, as given in Bodenheimer
and Pollack (1986). 
The assumption of hydrostatic equilibrium has been verified, at least
in the case of a constant accretion rate of solids, by Tajima and
Nakagawa (1997). 
The energy generation rate 
is calculated from the
accretion of  planetesimals and quasi-static
contraction. If planetesimals land on the core, the energy
deposition is smoothed over a region of about one core radius in
extent. 
The grain and molecular opacity in the
envelope is based on calculations by Pollack {\it et al.} (1985) and
Alexander and   Ferguson (1994), who used an interstellar size distribution.
Justification of this assumption and tests to determine the effect of a
large change in the opacity on the outcome of the calculations have
been discussed by Bodenheimer and Pollack (1986) and Pollack {\it et al.} 
(1996).
The equation of state is non-ideal and has been updated since 
the calculations of Pollack {\it et al.} (1996) were made; it is now 
based on calculations by Saumon {\it et al.} (1995), interpolated
to a near-protosolar composition of $X = 0.74, Y = 0.243, Z = 0.017$. 
 
The calculations are started at $t = 10^4$ yr\footnote\dag{Note that
the starting times used by Pollack {\it et al.} (1996) are not $t=0$, as
stated in the text, but 0.21 Myr for Jupiter models, 0.18 Myr for
Saturn models, and 1.25 Myr for the standard Uranus case U1.}  
with a core mass of 0.1 M$_\oplus$; the 
corresponding envelope mass is  $10^{-9} $ M$_\oplus$.
These initial values were chosen for computational convenience, 
and our results are completely insensitive to changes in the 
initial conditions. 

Boundary conditions, which are used for all cases described 
below,  are provided at the inner and outer edges of the
envelope. At the inner edge, the luminosity $L_r = 0 $ and the radius
$r = R_{core}$, where $R_{core}$ is determined from the current core
mass $M_{Z}$ and the core density $\rho_{core}$. The core is assumed
to have a uniform density and to be composed of rock, under the
assumption that ices have evaporated at the nebular positions considered
here. Thus, for the cases of 51 Peg b and $\rho$ CrB b,  $\rho_{core}$ = 6
g cm$^{-3}$ and for 47 UMa b $\rho_{core}$ = 5 g cm$^{-3}$. 

The outer boundary condition can be applied in three different ways,
depending on the evolutionary stage of the planet. During the first, or 
{\it nebular}, stage the outer edge of the planetary structure is in
direct contact with the nebula gas. The outer density $\rho_{neb}$ 
and temperature $T_{neb}$ are prescribed from nebula models and do 
not vary with time. The outer radius of the planet is assumed
to fall at a modified accretion radius $R_a$. Let the tidal, or Hill, 
radius be
$$  {R_H} = {a}{\biggl( { {M_p} \over {3 M_\star}}\biggr)^{1/3}}~~, \eqno(2) $$
where $a$ is the distance to the central star, $M_p$ is 
the planet's mass,  and $M_\star$ is the star's mass.
Then $R_a$ is given by 
$${R_a} = { { GM_p}\over {{c^2} + {{GM_p} \over {R_H}}}}~~, \eqno(3)$$
where $c$ is the sound speed in the nebula.
In the limits of large and small $R_H$, this expression reduces to the 
accretion radius and the tidal radius, respectively. 
The gas accretion rate is determined by the requirement that the outer
radius of the protoplanet be close to $R_a$, within a small tolerance.       
At every time step mass is added at the outer edge so that this 
requirement is satisfied.

  The limiting gas accretion rate is determined by the ability of the
nebula to supply gas at the required rate.  In a typical nebula the
mass transfer rate at a given radius caused, for example,  by 
viscous effects, is $3 \times 10^{-8}$ M$_\odot$/yr or 
$\approx 10^{-2}$ M$_\oplus$/yr.  Once the limiting
rate is reached, the planet contracts inside $R_a$, and the
evolution enters the {\it transition} stage.    Here  the
planet is still assumed to be in hydrostatic equilibrium, but gas
is accreting hydrodynamically onto it at near free-fall velocities. 
Thus, the luminosity of the protoplanet is given by the sum of two
contributions, its normal luminosity ($L_{contr}$) caused by contraction and
accretion of solids, plus a gas accretion luminosity given by
$$ {L_{acc}} = { G M_p \dot M_{XY,max} (1/R_p - 1/R_a)}~\eqno(4) $$ 
where $R_p$ is the planetary radius and $\dot M_{XY,max}$ is the limiting 
accretion rate. The density $\rho_s$ and temperature $T_s$ at $R_p$  
are also modified. The density  is obtained from 
$$  {\rho_0} =  {{ \dot M_{XY,max}} \over { 4 \pi R_p^2 v_{ff}}}~~,\eqno(5)$$
where $\rho_0$ is the density at the inner edge of the infalling flow
and $v_{ff}$ is the free-fall velocity from $R_a$ to $R_p$. Then
$\rho_0$ is multiplied by the square of the infall Mach number to
get the planetary boundary density $\rho_s$, under the assumption 
that the boundary shock is isothermal. To get the temperature, the
radiative diffusion equation 
$$ { {dT^4} \over {dr}} = - { { {3} \over {4 \sigma}} { {\kappa \rho_0 
R_p^{1.5}} \over {r^{1.5}}} { {L_r} \over {4 \pi r^2}}}~ \eqno(6)$$
is integrated and solved approximately under the assumptions that the
Rosseland mean opacity, $\kappa$, and the total luminosity,  
$L_r = L_{acc} + L_{contr}$, are constants as a function of distance $r$. 
In the limit that the envelope is optically thin the result is simply 
$T_s = T_{neb}$. These approximations have been checked against more
detailed solutions of the radiative transfer equation and found to 
be quite adequate. 

The supply of gas to the planet is eventually assumed to be 
exhausted ({\it e.g.}, as a result of tidal truncation of the
nebula, removal of the gas by effects of the star, and/or the accretion
of all nearby gas by the planet), and 
the planet's mass levels
off to some limiting value, defined by the planetary  system that is being
modeled (assuming sin $i \approx  1$,  except in Case P1a). 
This process is not modeled in detail, but $\dot M_{XY}$ onto the
planet is assumed to reduce smoothly to zero as the limiting value is
approached.  The planet then evolves through the {\it isolated} stage, 
during which it remains at constant mass. The boundary conditions
are then simply the standard photospheric conditions 
$$ L = 4 \pi \sigma_{SB} R_p^2 T_s^4~, \eqno(7)$$
where $\sigma_{SB}$ is the Stefan-Boltzmann constant, and 
$${ \kappa P} = {{2} \over {3}} { g}~~, \eqno(8)$$
where $P$ is the photospheric pressure and $g$ is the acceleration 
of gravity at $R_p$.  If the planet is close to the
central star, insolation is taken into account according to the 
procedure described by Stringfellow {\it et al.} (1990). 

\section{RESULTS}

\begin{table*}[!ht]
\centering
\resizebox{1\linewidth}{!}{%
\includegraphics[trim={2cm 7cm 0cm 2cm},clip]{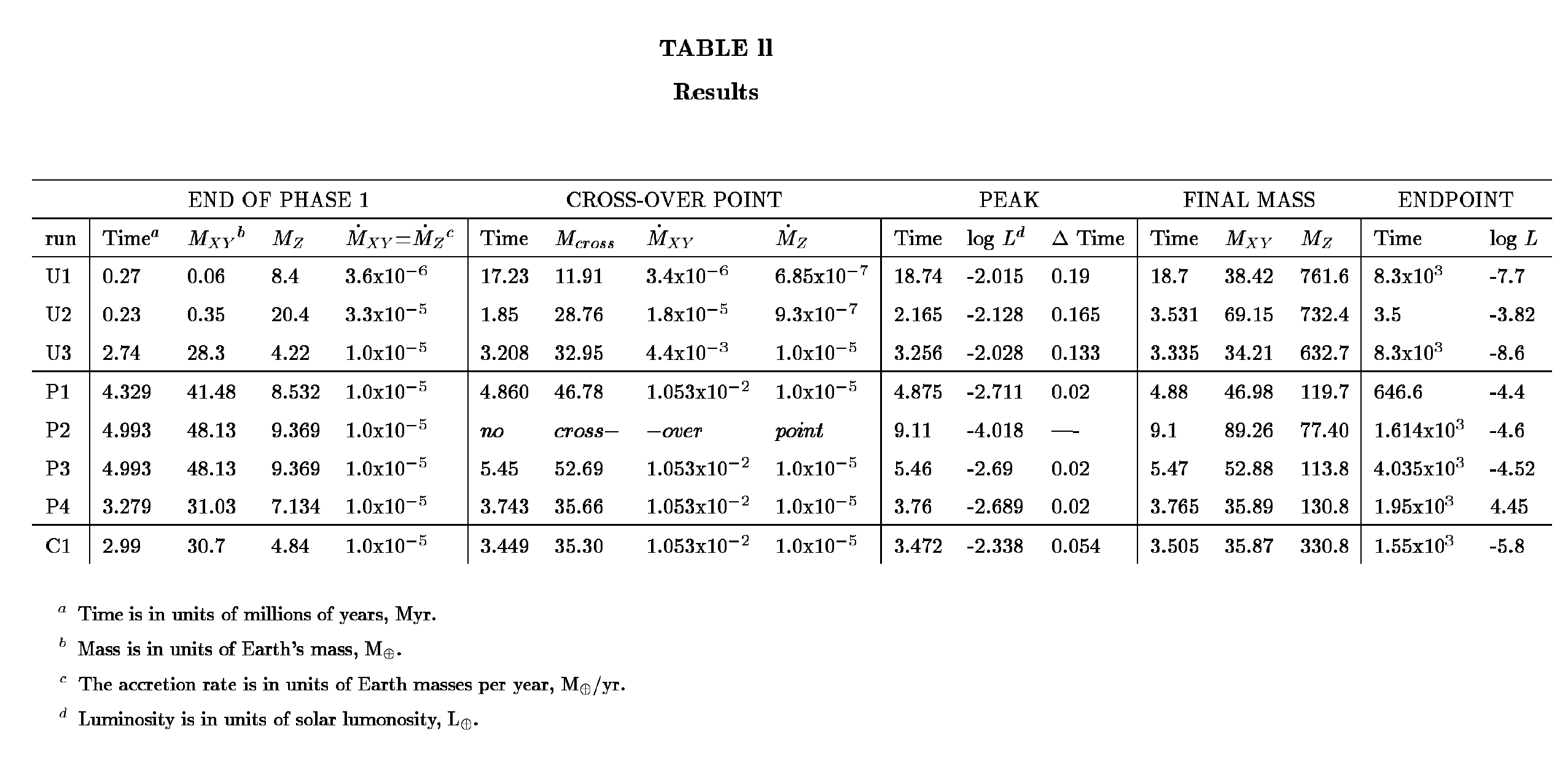}}
\end{table*}

\subsection{\emph{Formation of the companion to 47 UMa}}

The companion to 47 UMa, a solar-type star, has a nearly circular orbit, 
a period of 3 years,  an orbital radius of 
2.1 AU, and a minimum mass of 2.38 M$_J$ = 757 M$_\oplus$ 
(Marcy and Butler 1998). We perform two types of  calculations, the first      
(Cases U1 and U2) in which the solid accretion rate is 
calculated in detail according to the procedures
described in Pollack {\it et al.} (1996) and summarized in the previous
section, and the second (Case U3) in which the solid
accretion rate is assumed to have a  constant value
of $10^{-5}$ M$_\oplus$ yr$^{-1}$. 
The work of Bell {\it et al.} (1997)
provides detailed disk models for a range of values
of the assumed steady-state accretion rate of nebular mass ($\dot M_{neb}$) and
viscosity parameter ($\alpha$). These models are used to determine 
boundary conditions in the nebular stage. The results are summarized in 
Table II.   

For Cases U1 and U2  the crucial parameter is the assumed surface
density of solid material in the disk, $\sigma_{init,Z}$ (Pollack 
{\it et al.} 1996). In that paper the solid material was made up of
H$_2$O ice, rock, and CHON, with mass fractions 40\%,  30\%, and 
30\%, respectively. At 2.1 AU in all reasonable disk models the
ices have evaporated. A trial value of $\sigma_{init,Z} = 50 $ g cm$^{-2}$
is assumed for Case U1.  An examination
of the models of Bell {\it et al.} (1997) for $\alpha = 10^{-2}$ shows
that for the required corresponding total surface density, the
nebular temperature at 2.1 AU will be $\approx 1000$ K, too hot for 
the survival of the CHON material. Thus the gas-to-solid ratio will
be about 230, the total surface density  about $1.2 \times 10^4$ 
g cm$^{-2}$, and the gas density $\rho_{neb}$  
about $2 \times 10^{-9}$ g cm$^{-3}$. 
The corresponding Bell model has $\dot M_{neb} = 10^{-5}$ 
M$_\odot$/yr 
for $\alpha = 10^{-2}$, a mass transfer rate that is much higher 
than that deduced for typical disks around young stars. 
Correspondingly, the surface density is more than 10 times higher
than that in the ``minimum mass" solar nebula at the
same position (it is not known, of
course, whether that nebular model has any relevance to the case 
of 47 UMa). Even with the high surface density, the disk models
for Cases U1 and U2 are gravitationally stable at 2.1 AU, although
they may be gravitationally unstable at $\sim 20 $ AU (Boss 1996; 
Bell {\it et al.} 1997). 
The surface density of the solid 
material, which is in the form of rocky planetesimals of radius 100 km, 
is taken to be constant in space, and the decrease in
this surface density caused by accretion onto the planet is taken into
account. The ablation and evaporation of the planetesimals as they
travel through the gaseous envelope is calculated, but the evaporated
material, which has a higher molecular weight than the ambient gas,
is then assumed to settle slowly to the core boundary, releasing
gravitational energy in the process.   

As discussed in Pollack {\it et al.} (1996), the nebular stage of
the evolution is divided into three phases. During the first, which
lasts $1.0 \times 10^5$ yr in Case U1, 
the solid accretion rate increases rapidly, 
peaks, and then declines steeply as the planet approaches its 
isolation mass. The gas accretion rate during this phase is very
low, but it increases rapidly. The second phase involves gradual
accretion of gas and solids at nearly constant rates, with the gas
accretion rate exceeding that of the solids. This phase lasts 
$1.7 \times 10^7$ yr. When  $M_{cross}$ 
is approached, the third phase, rapid gas accretion, sets in.
The calculations reported by Pollack {\it et al.} (1996) were
stopped during this third phase; here they are continued into the
transition stage and the isolated stage.

\begin{figure*}
\centering
\resizebox{1.0\linewidth}{!}{%
\includegraphics[trim={1cm 1cm 6cm 6cm},clip]{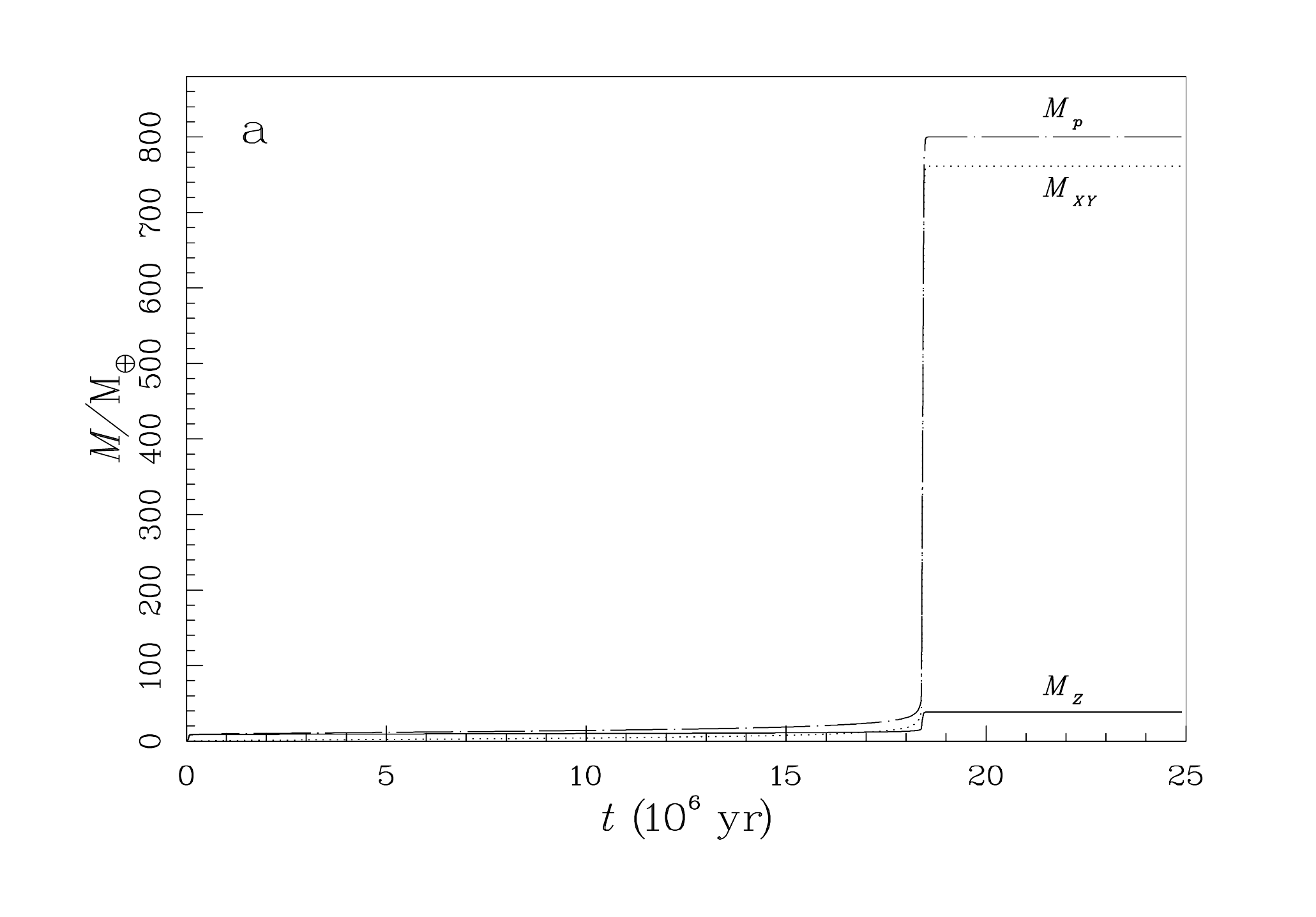}%
\includegraphics[trim={1cm 1cm 6cm 6cm},clip]{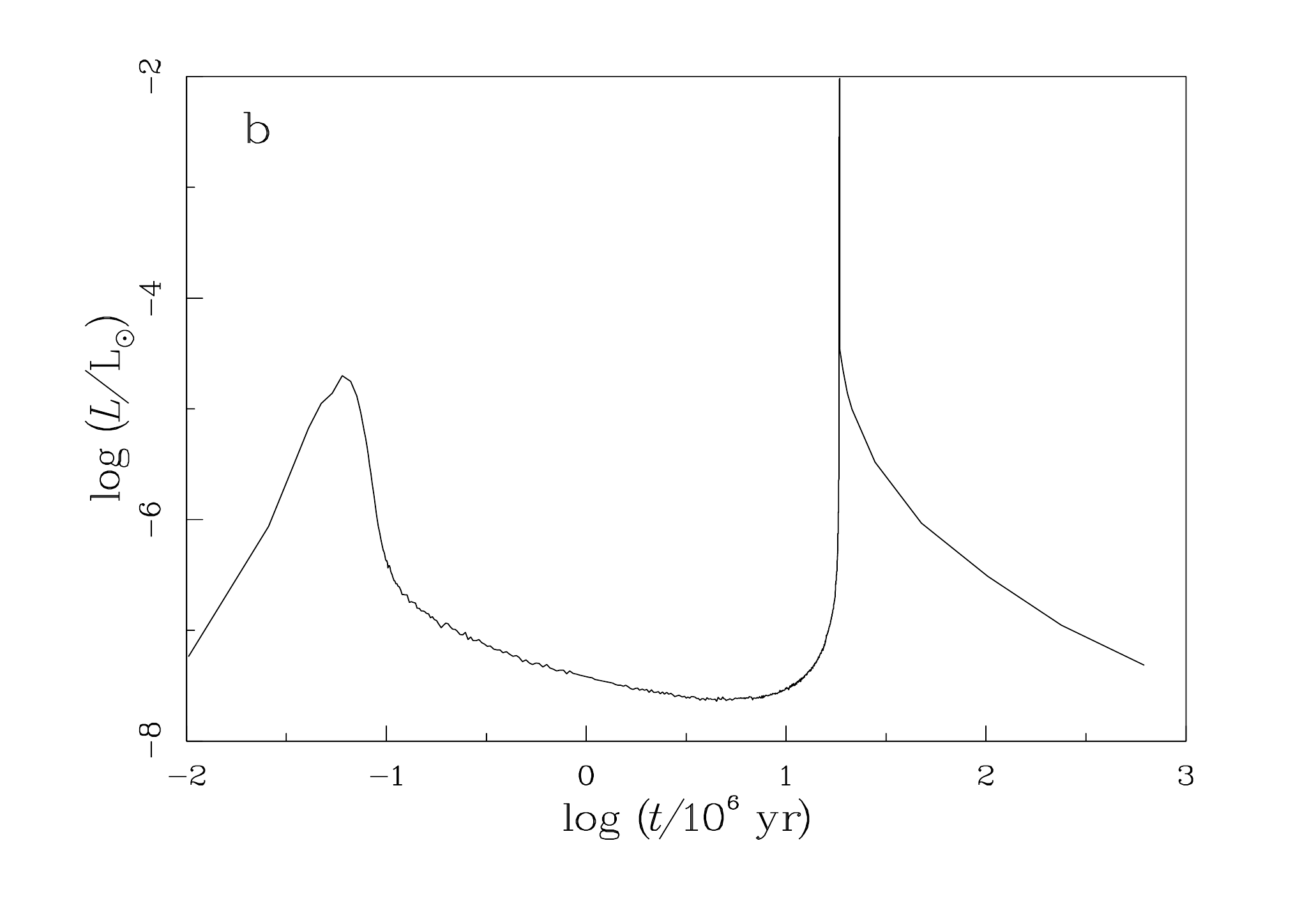}}
\resizebox{1.0\linewidth}{!}{%
\includegraphics[trim={1cm 1cm 6cm 7cm},clip]{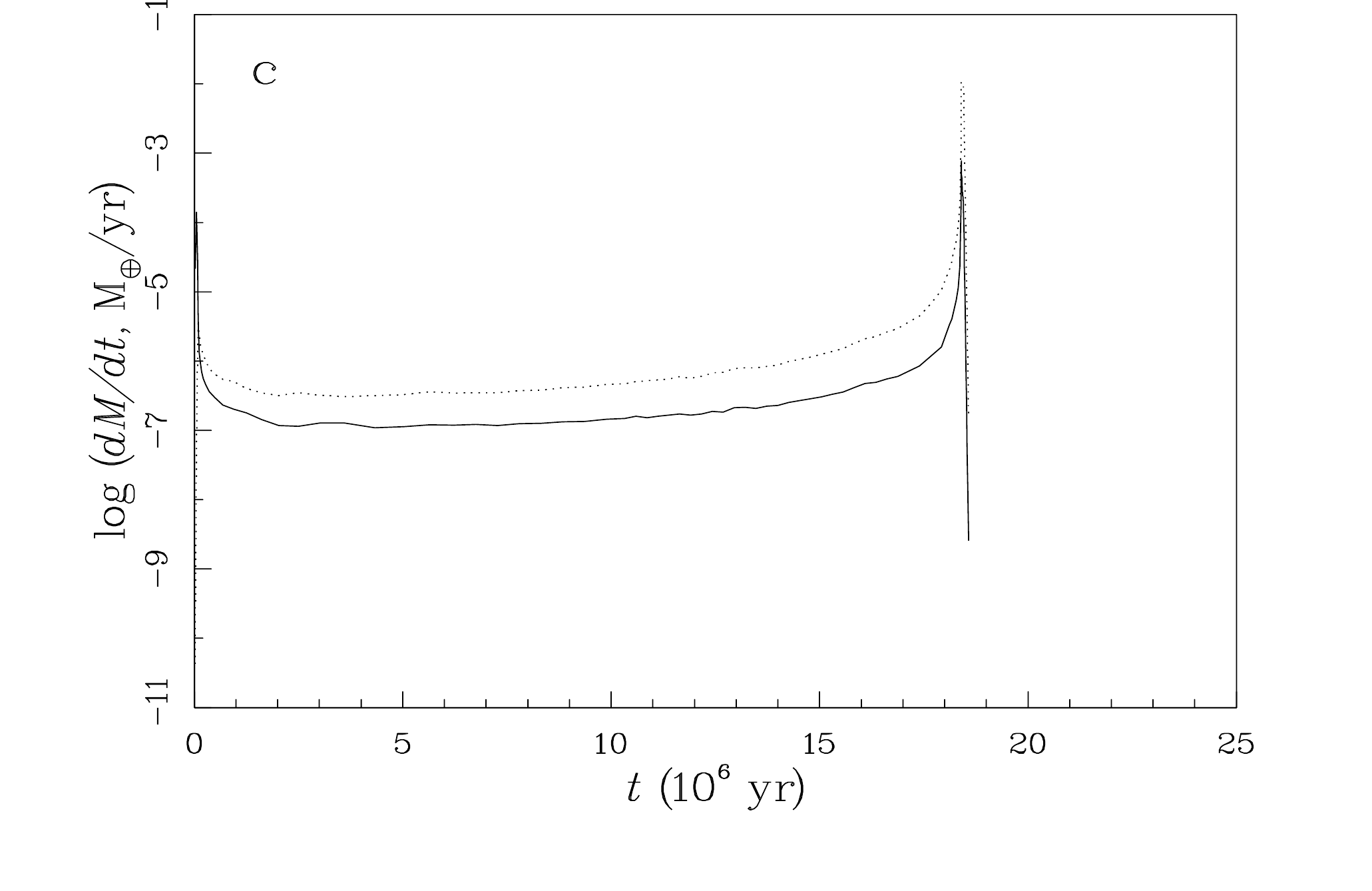}%
\includegraphics[trim={1cm 1cm 6cm 7cm},clip]{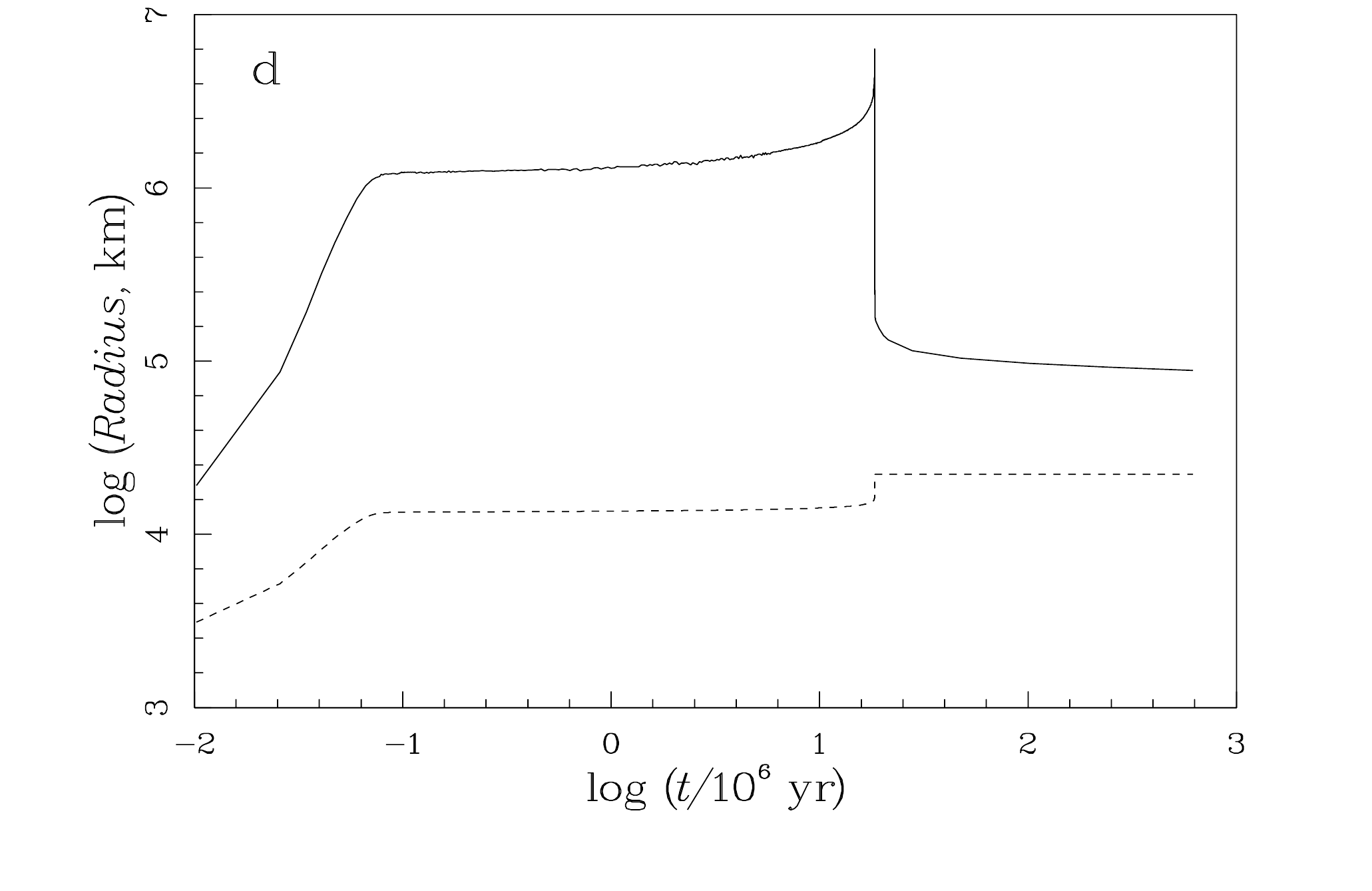}}
\caption{
The growth and subsequent 
evolution of the companion to 47 UMa     in a 
massive  protoplanetary disk with the solid accretion rate 
calculated as a function of time (Case U1). 
a) Mass of the solid component ({\it solid line}), 
the gaseous component ({\it dotted line}), 
and the total mass ({\it dot-dashed line}) as  functions of time. 
b) Luminosity radiated by the planet as a function of time. 
c) Logarithms of the mass accretion rate of planetesimals
({\it solid line}) and the accretion rate of gas ({\it dotted line}) 
as  functions of time. 
d)  Radii of the solid core 
({\it dashed line}) and the entire protoplanet ({\it solid line})
as  functions of time.
}
%\label{}
\end{figure*}

The results are shown in Fig. 1. During the first phase, at a 
time of $1.0 \times 10^5$ yr,  the
solid accretion rate peaks at 10$^{-4}$ M$_\oplus$~yr$^{-1}$ and the 
luminosity at $2 \times 10^{-5}$ L$_\odot$. The estimate of the 
isolation mass (Eq. 14 of Pollack {\it et al.} 1996) 
corresponding to the choice of $\sigma_{init,Z}$ and $a$ is 8.5 M$_\oplus$,
in excellent  agreement with the calculated value of 8.4 
M$_\oplus$ (Table II).  At the end of phase 1, the mass
in the gaseous envelope $M_{XY} =  6 \times 10^{-2}$
 M$_\oplus$.
As Figs. 1b and 1c show, the luminosity then decreases to a 
nearly constant value of $ 3 \times 10^{-8}$ L$_\odot$, and 
$\dot M_{Z}$ and $\dot M_{XY}$ are 10$^{-7}$ and 
$3 \times 10^{-7}$ M$_\oplus$~yr$^{-1}$,  respectively. These values
then increase rapidly as the phase of rapid gas accretion starts. The peak
luminosity reaches 10$^{-2}$ L$_\odot$, and the width of the peak,
measured at log ($L/{\rm L_\odot}) = -4.5$, is 
$1.9 \times 10^5$ yr.  Note that the heights and widths of the 
luminosity peaks quoted in this paper are at best order of magnitude
estimates because of our {\it ad hoc} treatment of the limiting gas
accretion rate and the termination of accretion, but the product 
$\int L dt$ is a more robust result.
The evolution time to the beginning of the isolated
phase is $1.8 \times 10^7$ yr, longer than reasonable 
nebular lifetimes. The final values of $M_{Z}$  and
$M_{XY}$  are 38.4 and 761.6 
 M$_\oplus$, respectively, giving a total mass of 2.52 M$_J$ = 800 M$_{\oplus}$. 

\begin{figure}
\centering
\resizebox{1.0\linewidth}{!}{%
\includegraphics[trim={1cm 1cm 6cm 6cm},clip]{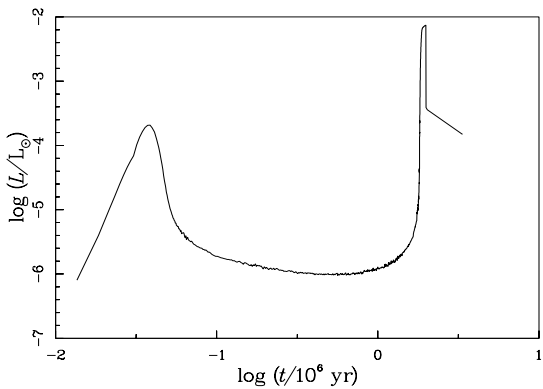}}
\caption{
The luminosity radiated by the planet
as a function of time for Case U2.
}
%\label{}
\end{figure}

The results of Pollack {\it et al.} (1996) for models of this type show that
the formation time scale, which is determined primarily by the duration
of phase 2, is very sensitive to $\sigma_{init,Z}$. 
Case U2 was run with the same parameters as Case U1 but with $\sigma_{init,Z}$
increased from 50 to 90 g cm$^{-2}$. The quantity $T_{neb}$ was increased
to 1100 K at the same time, but that should have very little influence
on the results (see below).  The results show that, compared with
case U1, 
the isolation mass is increased by more than a factor 2 in phase 1, and 
$\dot M_{XY}$ and $\dot M_{Z}$ increase by more than a factor 10 in phase 2. 
The rapid gas accretion occurs in a manner similar to that in case U1,
with a peak of 10$^{-2}$ L$_\odot$ and a width
(measured at log ($L/{\rm L_\odot}) = -3.4$) of $1.65 \times 10^5$ yr
(Fig. 2). The final values of $M_{Z}$  and
$M_{XY}$  are 69 and 732 M$_\oplus$, respectively, and the planet
forms in just over $1.9 \times 10^6$ yr, about a tenth of the time
needed for Case U1. 

\begin{figure*}
\centering
\resizebox{1.0\linewidth}{!}{%
\includegraphics[trim={1cm 1cm 6cm 6cm},clip]{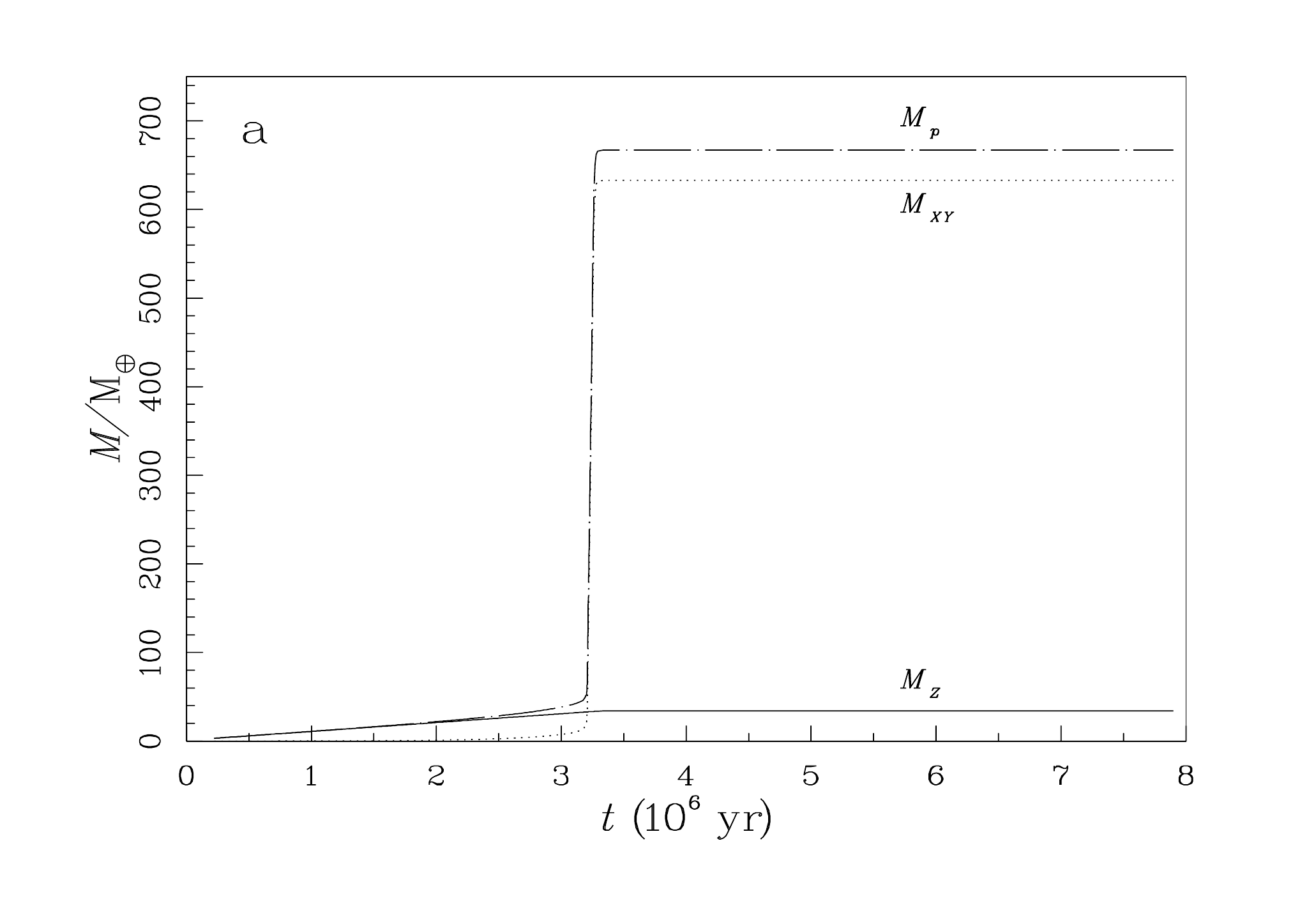}%
\includegraphics[trim={1cm 1cm 6cm 6cm},clip]{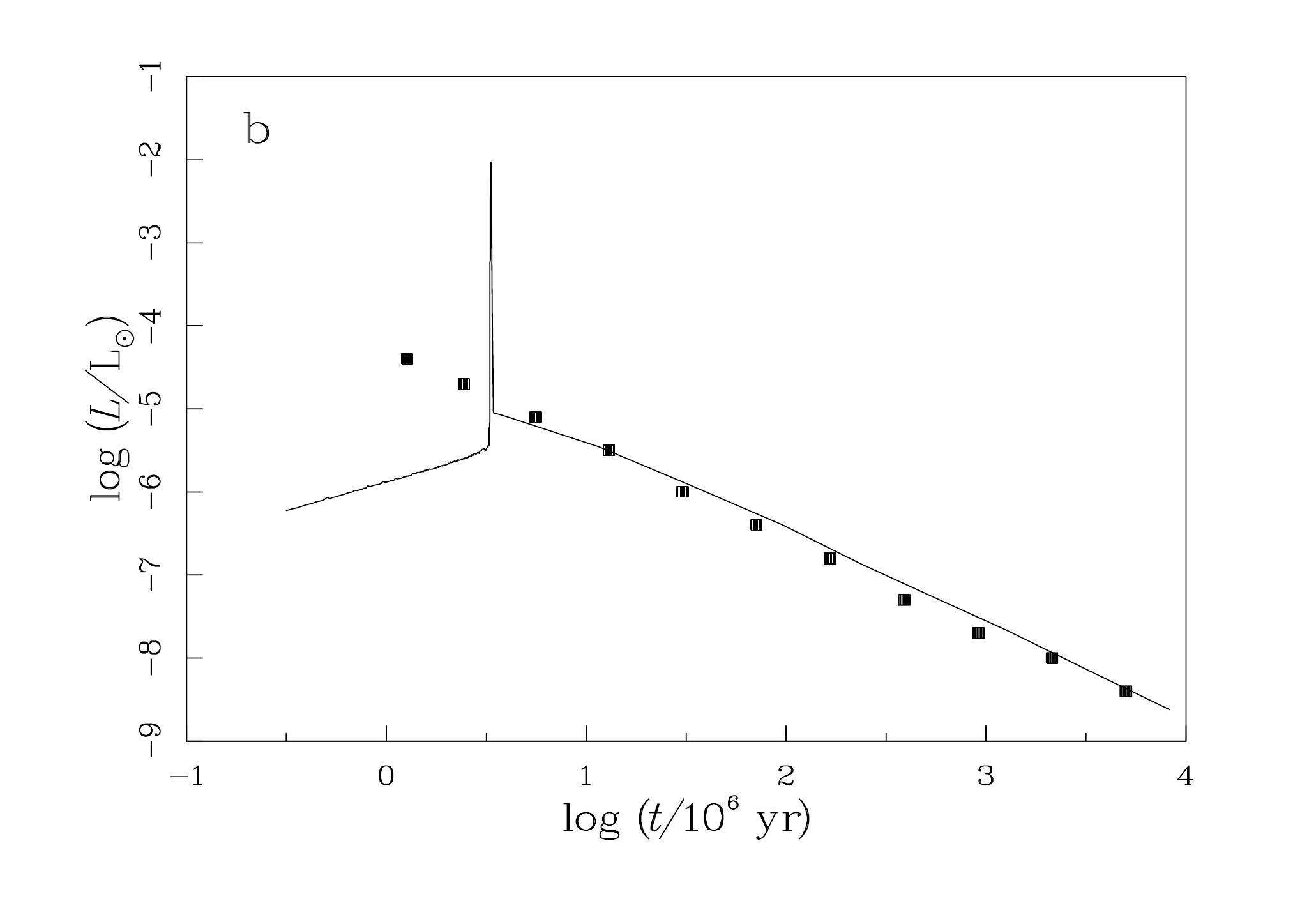}}
\caption{
The evolution of the companion to 47 UMa     in a 
standard protoplanetary disk and with constant solid accretion rate 
(Case U3). a) Mass of the solid component, 
the gaseous component, and the total mass as functions of time. 
b) Luminosity as a function of time. The solid squares        
represent the solution of Burrows {\it et al.} (1997) for an
isolated planet of 2.0 M$_J$.
}
%\label{}
\end{figure*}

The results of Case U1 and U2  show that the evolution can be reasonably
approximated by an assumed constant rate of solid accretion 
throughout much of phase 2. Thus for case U3 we assume that $\dot M_Z$
is constant at $10^{-5}$ M$_\oplus$ yr$^{-1}$  throughout the entire
evolution, a rate slightly less than the mean accretion rate for Case U2. 
An accretion rate of this order of magnitude is needed to obtain a 
reasonable formation time. 
The parameters for the nebular model for Case U3 are taken to be  
$\alpha = 0.01$, $\dot M_{neb} = 10^{-7}$ M\sol~~yr$^{-1}$. At 2.1 AU the
model gives $T_{neb}$ = 400 K and $\rho_{neb} =  3 \times 10^{-10}$ 
g cm$^{-3}$, corresponding to a total surface density
of about 1000 g cm$^{-2}$. The tidal truncation limit for these conditions is
2.3 M$_J$, very close to the observed  minimum mass. The masses as a 
function of time are plotted in Fig. 3a; a formation time of 3.4 
million years is obtained and the final values of $M_{Z}$ 
and $M_{XY}$ are 34 and 633 M$_\oplus$, 
respectively.  The radiative envelope in these  models contains a 
significant mass fraction; thus the analysis
of Stevenson (1982),
which assumes a constant $\dot M_Z$, 
should apply,  and the final core mass should be 
insensitive to the nebular boundary conditions. 
The luminosity as a function of time 
is plotted in Fig. 3b. The peak luminosity reaches 10$^{-2}$ L$_\odot$, 
essentially the same as that in Case U1, 
and the full width of the peak, corresponding to $L > 10^{-5}$ 
L$_\odot$, is $1.3 \times 10^5$ yr. 
The isolated phase is calculated
up to a time beyond 10$^9$ yr; the late-phase luminosity decline
is exponential. Insolation is not included in this case and the total
mass is cut off at 2.1 M$_J$ to facilitate a comparison with an
isolated protoplanet of  2 M$_J$ calculated 
by Burrows {\it et al.} (1997). As shown in Fig. 3b, the agreement
is very good.

\subsection{\emph{Formation of 51 Peg b}}

The companion to 51 Peg has an orbital period of 4.23 days, 
an orbital separation of 0.05 AU from the star,  and a 
minimum mass of 0.44 M$_J$ = 140 M$_\oplus$ (Marcy and Butler 1998). 
The first calculation presented here
assumes that the planet formed in a standard nebular disk. 
The  Bell {\it et al.} (1997) model for 
$\dot M_{neb}$ = 
10$^{-8}$ M$_\odot$ yr$^{-1}$
and $\alpha = 0.01$ gives a
midplane temperature in the disk at 0.05 AU of  about
1500 K, so some solid material could survive. We assume that the solid
material is supplied to the inner region of the disk by viscous evolution;
the above value for $\dot M_{neb}$ 
corresponds to $\approx 10^{-5}$ M$_\oplus$ yr$^{-1}$
of solid particles that can exist at 1500 K.  From the models, $\rho_{neb} = 
5 \times 10^{-8}$ g cm$^{-3}$ and $T_{neb}$ = 1500 K. All of the solid
material that arrives at 0.05 AU is assumed to be collected by the planet. 
The tidal truncation conditions (Lin and Papaloizou 1993), which give the
approximate maximum mass a giant planet can accrete before complete
gap formation, are given by
$$ M_{p,max} = {\rm max} [{{40 \nu}\over  {(\Omega a^2)} } {M_\star},~~
 {3} \left({ {c} \over {\Omega a}}\right)^3 { M_\star}].~\eqno(9) $$
Here $\nu$ is the nebular viscosity.
In the particular nebular model chosen, the viscosity condition  determines the
maximum mass, which is about 0.7 M$_J$. In the calculation, denoted 
as Case P1, the mass is
assumed to level off at 0.52  M$_J$ = 167 M\earth. 

\begin{figure*}
\centering
\resizebox{1.0\linewidth}{!}{%
\includegraphics[trim={1cm 1cm 6cm 6cm},clip]{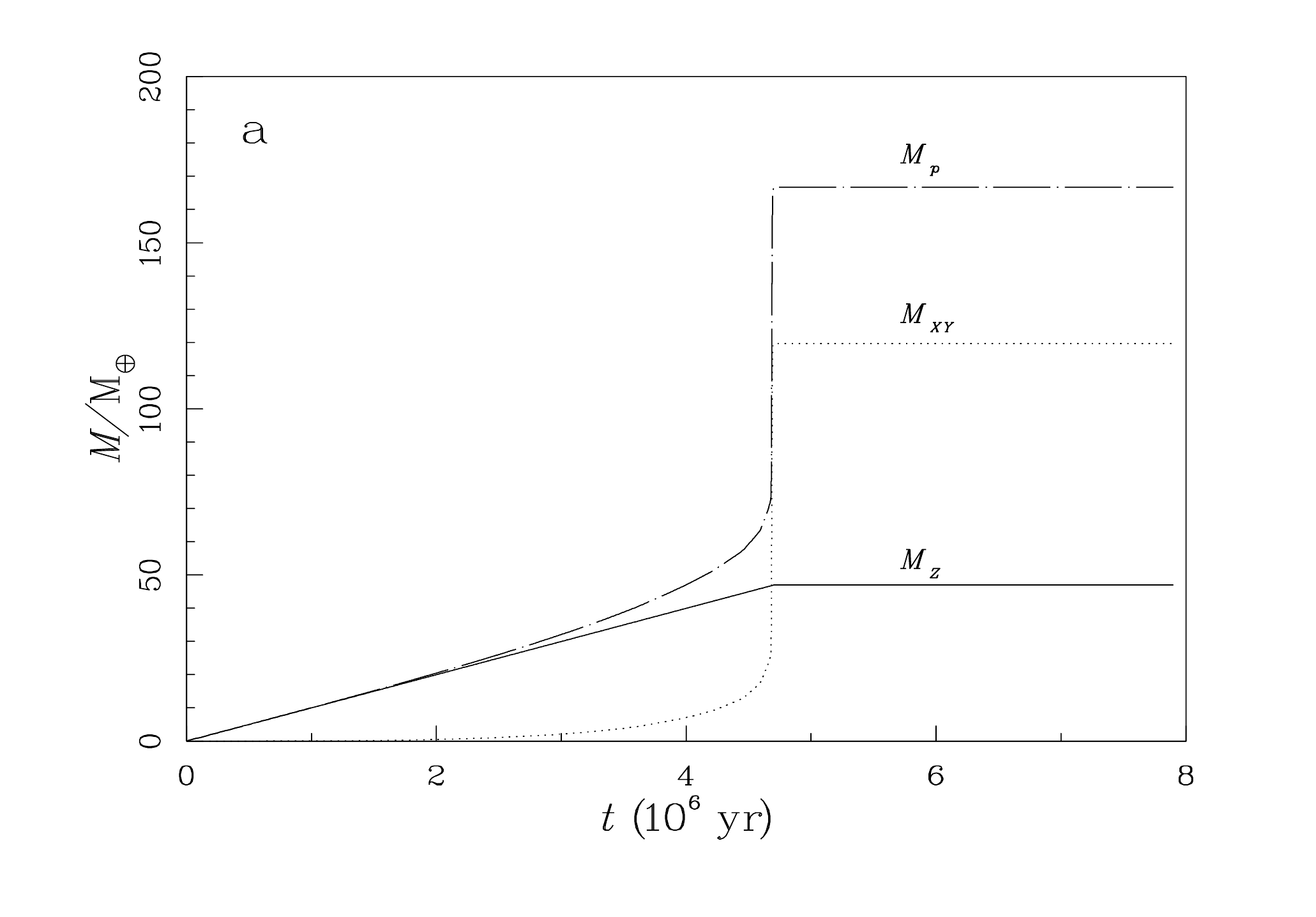}%
\includegraphics[trim={1cm 1cm 6cm 6cm},clip]{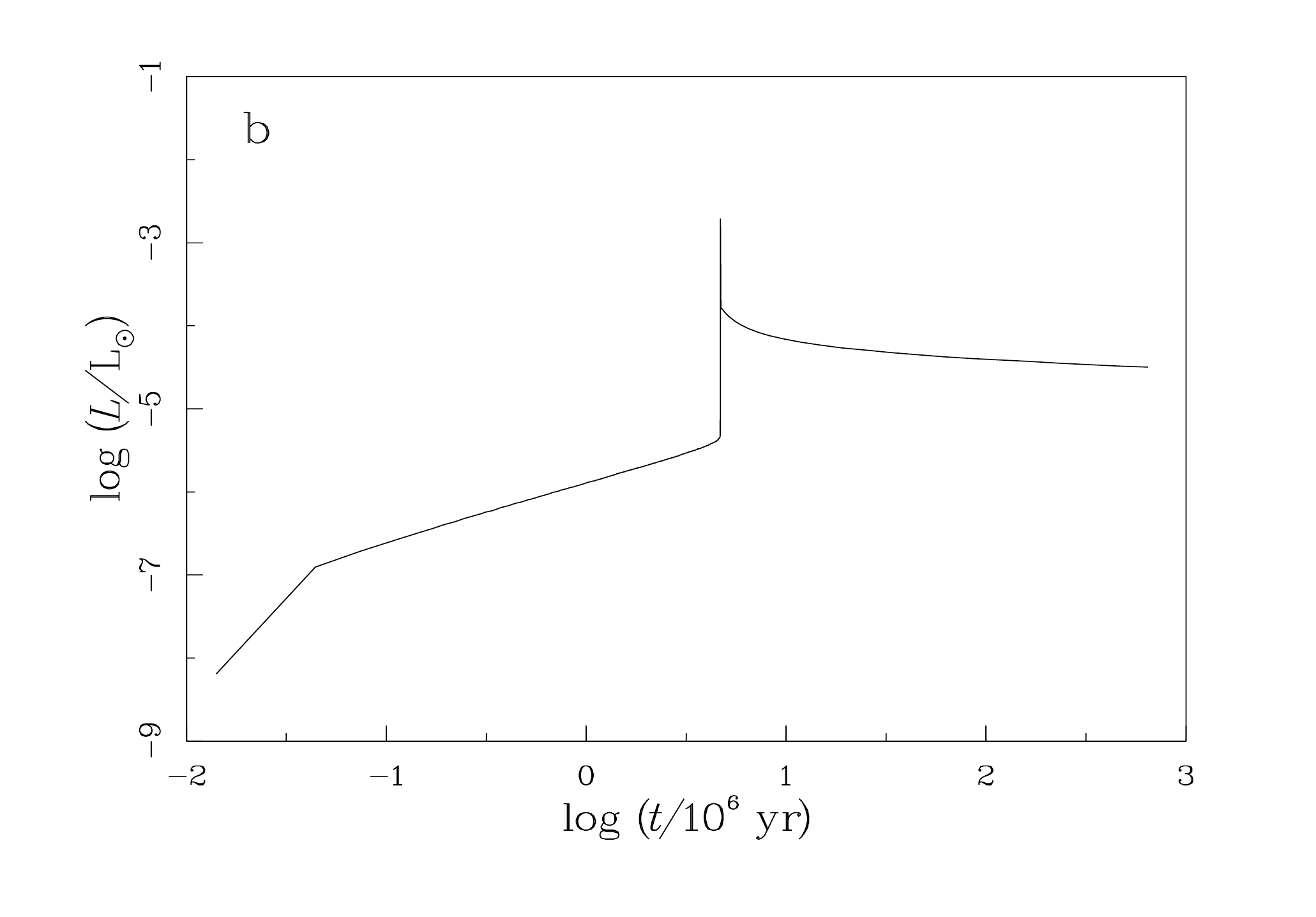}}
\resizebox{1.0\linewidth}{!}{%
\includegraphics[trim={1cm 1cm 6cm 7cm},clip]{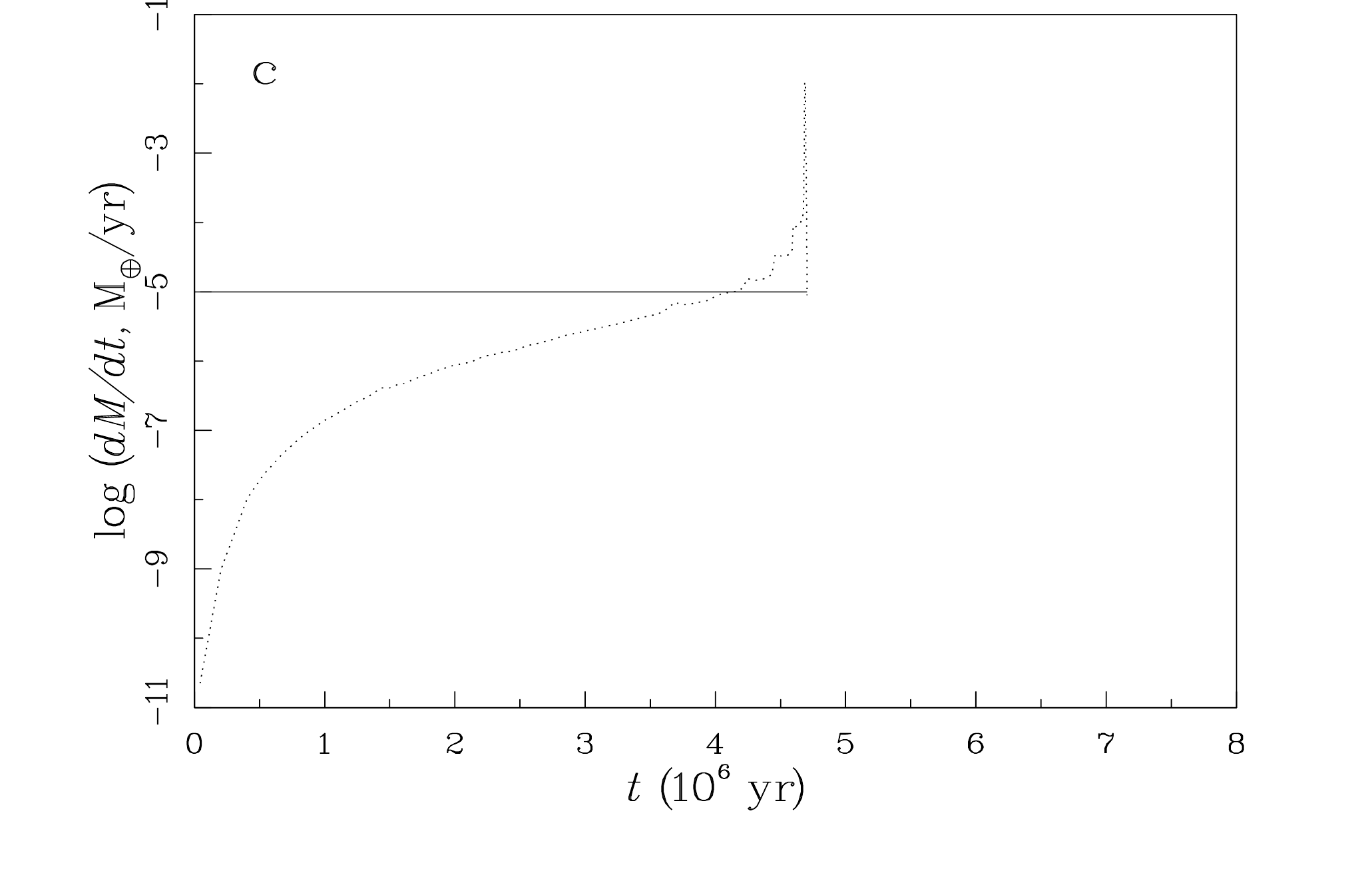}%
\includegraphics[trim={1cm 1cm 6cm 7cm},clip]{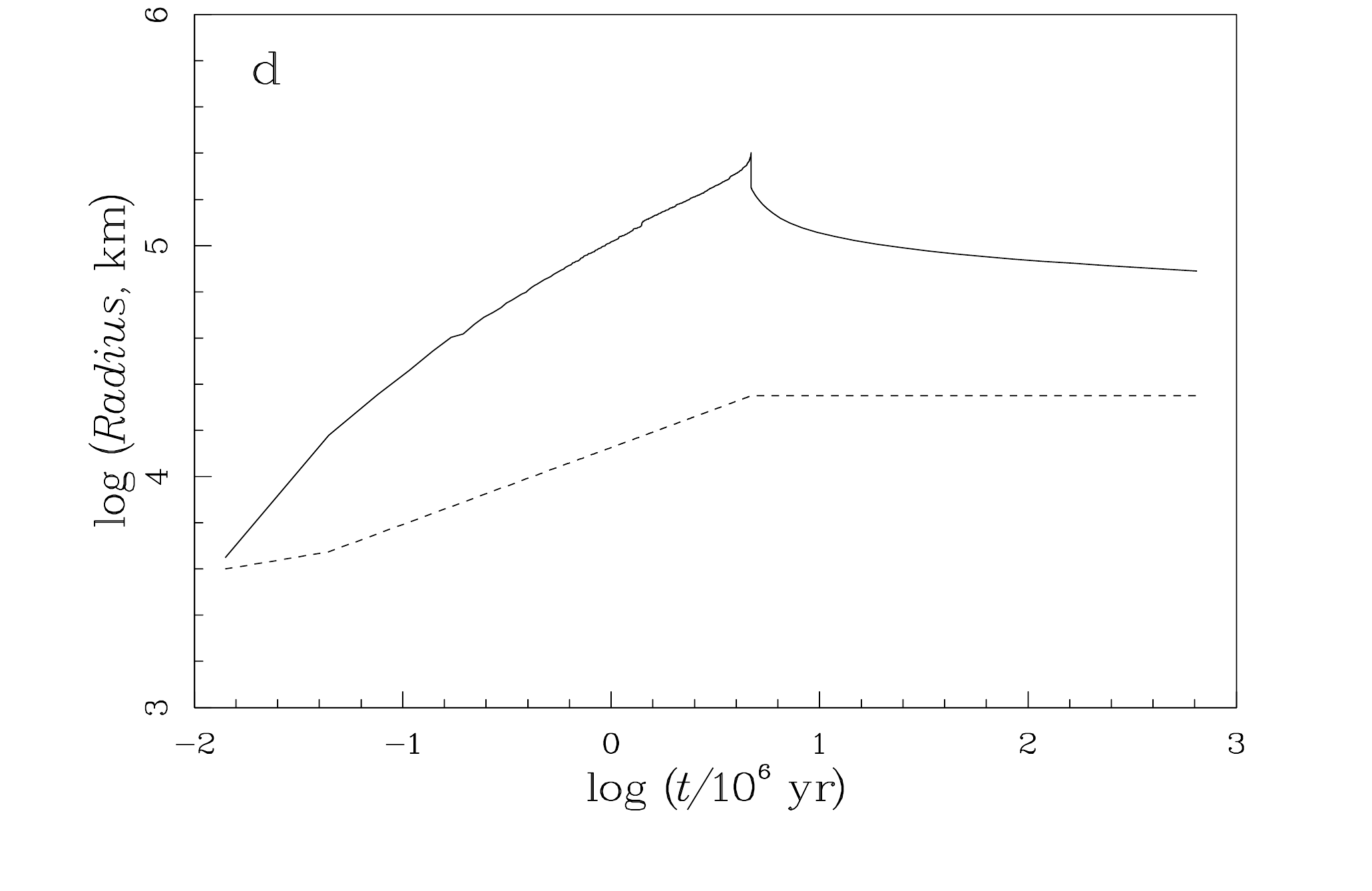}}
\caption{
The evolution of the companion to 51 Peg in a standard
protoplanetary disk (Case P1). a)  Mass of the solid 
component of the planet ({\it solid line}), the gaseous component 
({\it dotted line}), and 
the total mass ({\it dot--dashed line}) as  functions of time.
b) Luminosity of the protoplanet as a function of time.
The luminosity spike corresponds to the brief period of rapid
contraction of the planet inside the Hill radius and the hydrodynamic
accretion of nebular gas onto it. After about 10$^8$ yr the luminosity
levels off to log $(L/{\rm L}_\odot) = -4.5.$ 
c) Logarithm of the mass accretion rate of planetesimals
({\it solid line}) and the accretion rate of gas ({\it dotted line}) 
as  functions of time. The upper limit on the gas 
accretion rate is assumed to be 10$^{-2}$ M$_\oplus$ yr$^{-1}$. This
rate is assumed to drop smoothly to zero as the planet approaches
its final mass of 0.52 M$_J$.  d) Radii of the solid core 
({\it dashed line}) and the entire protoplanet ({\it solid line})
as  functions of time. The outer radius corresponds very closely to
that of the Hill sphere at times prior to the maximum. 
}
%\label{}
\end{figure*}

\begin{figure}
\centering
\resizebox{1.0\linewidth}{!}{%
\includegraphics[trim={1cm 1cm 6cm 6cm},clip]{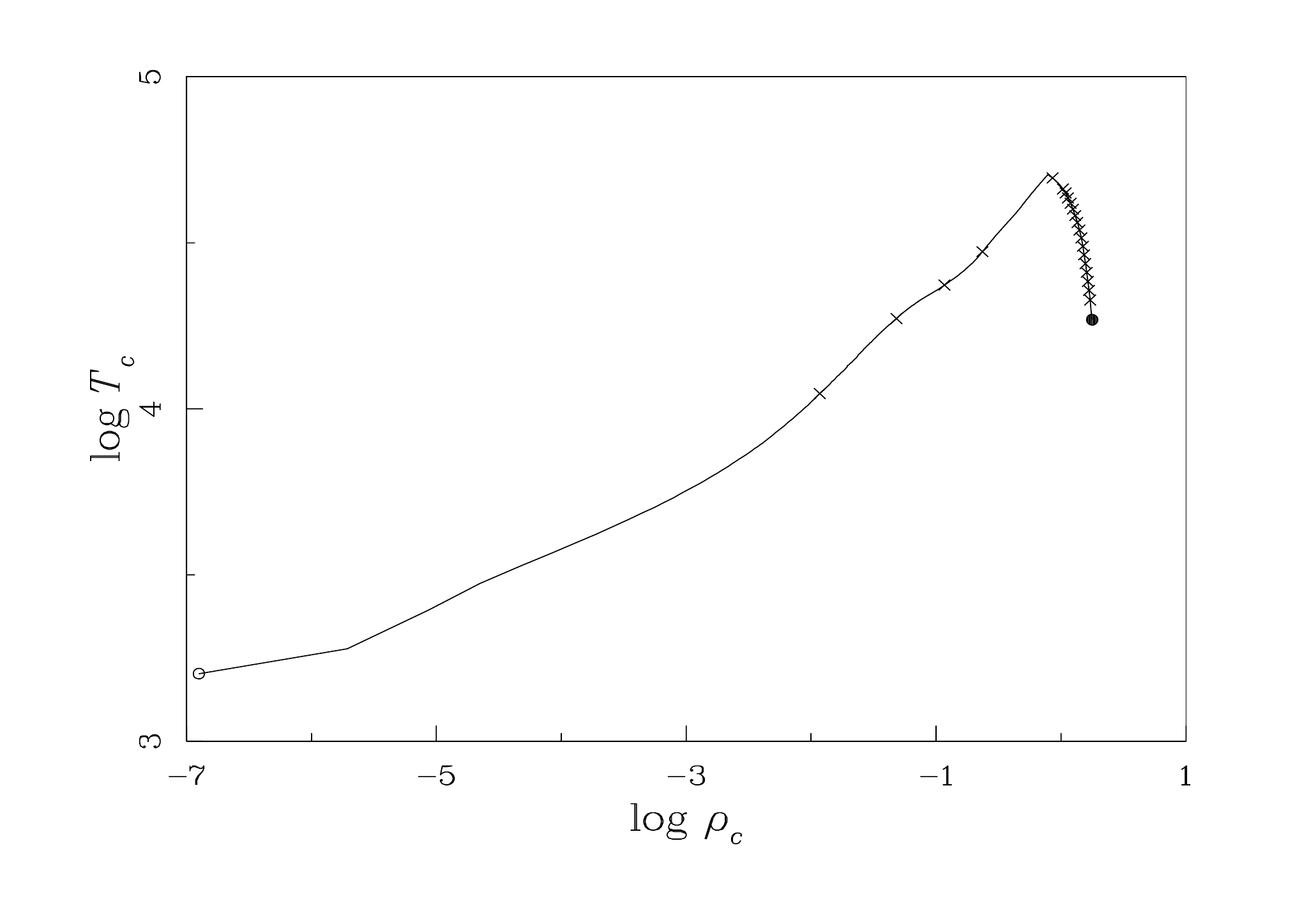}}
\caption{
The temperature of the gas    
as a function of density at the core/envelope boundary 
in Case P1. Marks on the curve
represent times, the first mark at  $10^6$ yr and at intervals
of  $ 10^6$ yr, ending at $2 \times 10^7$ yr. 
The decline in temperature begins  at the beginning of the 
isolation phase, when the planet
attains its final mass and gas accretion is shut off.
Open and closed circles correspond to the beginning and end of
the calculation, respectively. 
}
%\label{}
\end{figure}

The results (Figs. 4 and 5) 
show that the final mass of the solid core is 47 M$_\oplus$,
that of the gaseous envelope is 120 M$_\oplus$, and the time to reach final
mass is $4.70 \times 10^6$  years.
Figs. 4a,b,c,d  show the evolution of the mass of the core and envelope, 
the luminosity, the mass accretion rate, and the radius, respectively. 
During the early  evolution, the envelope mass is very
low compared with the core mass; only after $4.1 \times 10^6$ yr does the
envelope mass increase rapidly. The crossover mass ($M_Z = M_{XY}$)
is almost identical to the final  $M_Z$ and is reached 
at $4.7 \times 10^6$ yr. The transition stage, when
the planet first contracts within $R_a$, begins at about the same time.    

The luminosity at early times is dominated by accretion
of planetesimals. Between 4.68 and  4.7  Myr, when $ M_{XY}$ is 
comparable to $M_Z$, it becomes dominated by gravitational
contraction as a result of rapid gas accretion. However, because 
material is added at a surface temperature of 1500 K, the molecular
hydrogen dissociation zone lies fairly close to the surface. As the
material which is added contracts and is heated to higher temperatures, 
it undergoes dissociation in the range 3000 -- 6000 K. When  the gas
accretion rate becomes high, more than 90\% of the contraction energy
released by the entire mass must supply the dissociation energy of
the added mass. Thus, the emerging luminosity remains about constant 
at $10^{-5.5}$ L\sol. The rapid increase in $L$ corresponds to the 
transition stage and is almost entirely the accretion luminosity
liberated at the planetary surface as material falls from the Roche
radius to the surface. After a sharp spike, with maximum 
log $(L/{\rm L_\odot}) = -2.7  $ when  $M_{XY} = 119.6 $ M$_\oplus$, 
the accretion rate decreases as the planet reaches its terminal mass.
The width of the narrow spike in luminosity (Fig. 4b)
is only $2 \times 10^4$ yr, narrower than that in Cases U1, U2, and U3 
 because of the smaller amount of mass that is accreted. 
During the isolated phase ($t > 4.7 \times 10^6$ yr), 
the luminosity is at first produced by contraction and cooling 
at constant mass, but it 
soon becomes dominated by the heating of the planetary surface by the
star, and it levels off at log {\it L}/L$_\odot = -4.5$ for $t >  10^9$ yr.

The radius (Fig. 4d)  does not undergo 
major variation during the evolution. During
much of the nebular phase it is close to the Hill radius of 0.2 -- 0.3 R\sol,
reaching a maximum of 0.35 R\sol~at the transition stage, which is
indicated by the sudden drop in Fig. 4d. 
At the beginning of the isolated stage, 
the radius is 0.25 R\sol, small enough so that even in the presence
of the radiation of the nearby star the Jeans escape rate of material
from the outer atmosphere should be negligible. 
During the isolated stage, the radius decreases slowly to a final value of 
0.11 R\sol ~(approximately  10\% larger than Jupiter's present radius) 
after several Gyr. The observable surface temperature is 
$T_{neb}$ = 1500 K 
during the nebular and transition phases, and during the isolated
phase it soon levels off to the insolation value of 1300 K. The 
evolution of the interior is illustrated in Fig. 5. Early in the     
calculation, when  $M_Z$ = 1 M\earth, the central 
temperature $T_c \approx 2000 $ K  and the central density $\rho_c \approx
3 \times 10^{-6}$ g cm$^{-3}$. 
At the end of the transition stage $T_c$ reaches
a maximum of $4.85 \times 10^4$ K. Thereafter, it declines slowly
as the internal luminosity is supplied mainly by cooling of the interior
rather than by contraction. After 5 Gyr, $\rho_c = 2$ g cm$^{-3}$ 
and $T_c = 1.2 \times 10^4$ K. During the entire evolution,  energy 
transport is by convection through more than 99\% of the mass. A thin
surface radiative zone exists for $T < 2300$ K.  

The calculation was repeated (Case P1a) 
with a final assumed mass of 1.05 M$_J$ = 333 M\earth. 
The final value of the core mass was 47.27 M\earth, only slightly larger than that
for Case P1, and the envelope mass was 286 M\earth, a factor of 2.4 larger. 
The luminosity curve was essentially the 
same as that shown in Fig. 4b, but  the width of the spike at
4.7 Myr was increased to 4.9 $\times 10^4$ yr, a factor of 2.45 larger than
in Case P1, because of the accretion of more matter at the same limiting rate
as in that case.  The peak luminosity also increased by a factor 2.7, because 
of the higher value of $M_p$ in the calculation of the  accretion luminosity 
(Eq. 4). 
 
Another test was performed in which the uncertain grain 
opacity in the range 1500 $< T <$ 1800 K  was reduced 
by about a factor 10 (Case P1op). Otherwise the parameters were the 
same as in Case P1. The final
value of $M_Z$ was changed (downwards) by only about 3\%, and the peak
luminosity was slightly reduced (Table II). 
The grains occupy only a thin surface layer
in this model, so evidently it is the deeper molecular layers, just
above the convection zone, that are important in determining the
luminosity and thereby the evolutionary time scale to reach $M_{cross}$.
 
Models of star formation theory suggest that 
the stellar dipole magnetic field truncates the
disk out to about 0.1 AU (K\"onigl 1991; Shu {\it et al.} 1994; Ward 1997a)
and channels the flow of matter from disk to star. 
A  further calculation (Case P2)  was  therefore made with a
highly reduced disk density in the vicinity of the protoplanet.
If the planet is
interior to the 2:1 resonance with the inner edge of the disk, the
tidal effect is reduced by several orders of magnitude, so planetary
migration is less likely than in Case P1.
Chunks of rock can, however, migrate through the
disk and assemble into a protoplanet at 0.05 AU (Ward 1997a). We assume
that a small amount of gas still remains in the cavity. 
In this case we take $\rho_{neb} = 
5 \times 10^{-11}$ g cm$^{-3}$, a factor of 1000 smaller than that
in the previous case. Otherwise, the calculation is identical to Case P1. 
The values of $M_{Z}$ and $M_{XY}$ turn out to be
52.88 M\earth~and  113.8 M\earth, respectively, 
yielding a total mass of 0.52 M$_J$, the same as in Case P1.
The formation time to final
mass is $5.29 \times 10^6$ yr. 
Evolution of the radius, $T_{eff}$, $L$,  and the interior conditions are very
similar to that in the previous case. The close similarity of the core
masses in Cases P1 and P2 indicate that the conclusions of 
Mizuno (1980) and  Stevenson (1982) -- namely that 
the value of $M_{cross}$  is  insensitive to the outer   boundary
condition -- still hold even in this case of a protoplanet with relatively 
small radius. 
The small difference that was found is a result of the fact that the    
present models are largely convective. 
Wuchterl (1993) has emphasized the
fact that in this case
the core and envelope masses do depend on the outer boundary condition.

A further test was made (Case P3) which was identical to Case P2 except
that the limiting gas accretion rate was also reduced, to a value of
$2 \times 10^{-5}$ M\earth~yr$^{-1}$. In this case the crossover
mass was never reached, and the final values of $M_Z$ and 
$M_{XY}$ were 89.3 and 77.4 M\earth, respectively, with the total mass the same 
as that for Case P1. The formation time
to final mass was $8.9 \times 10^6$ yr. Because of the low limiting
gas accretion rate, the maximum value of the luminosity was 10$^{-4}$
L$_\odot$, more than a factor 10 less than that in the previous case. 

Case P4 is identical to Case P1 except that $\rho_{neb}$ is set to 
$5 \times 10^{-6}$
g cm$^{-3}$. The limiting gas accretion rate is also the same. 
Except for relatively small adjustments in
the core and envelope mass, which turn out to be 36 and 131 M\earth,
respectively, the results are quite similar to those of Case P1. This 
result is significantly different from that reported by Wuchterl (1997), 
who uses the same value of $\rho_{neb}$ 
but finds that rapid gas accretion starts
with a core mass of 13.5  M\earth.  However his $T_{neb} = 1252$ K  is lower
than ours, and his solid accretion rate is not constant but is 
calculated according to a variant of Eq. (1). 

\subsection{\emph{Formation of the companion to $ \rho$ CrB}} 

The solar-type star $ \rho$ CrB has a low-mass companion with orbital
period of 40 days, distance 0.23 AU, orbital eccentricity consistent 
with zero, and a minimum mass of 1.1 M$_J$ = 350 M\earth. To model the formation of
the planet (Case C1), we take nebular conditions from the Bell {\it et al.} 
(1997) 
model with viscosity parameter $\alpha = 0.01$ and 
$\dot M_{neb}  = 10^{-7}$ M\sol~
yr$^{-1}$. At 0.23 AU we have $T_{neb}$ = 1200 K and $\rho_{neb} =  5 \times 
10^{-8}$ g cm$^{-3}$. Again, in this region of a disk there is
insufficient mass in the neighborhood of the orbit to supply a Jupiter
mass, so we assume that solid material migrates into the region 
and collects on the protoplanet at a rate
of 10$^{-5}$ M\earth~yr$^{-1}$. The maximum rate of gas accretion is
taken to be 10$^{-2}$ M\earth~yr$^{-1}$. The tidal truncation limit 
for these nebular conditions (Eq. 9) is 1.4 M$_J$, consistent
with the probable mass of the planet. 

\begin{figure*}
\centering
\resizebox{1.0\linewidth}{!}{%
\includegraphics[trim={1cm 1cm 6cm 6cm},clip]{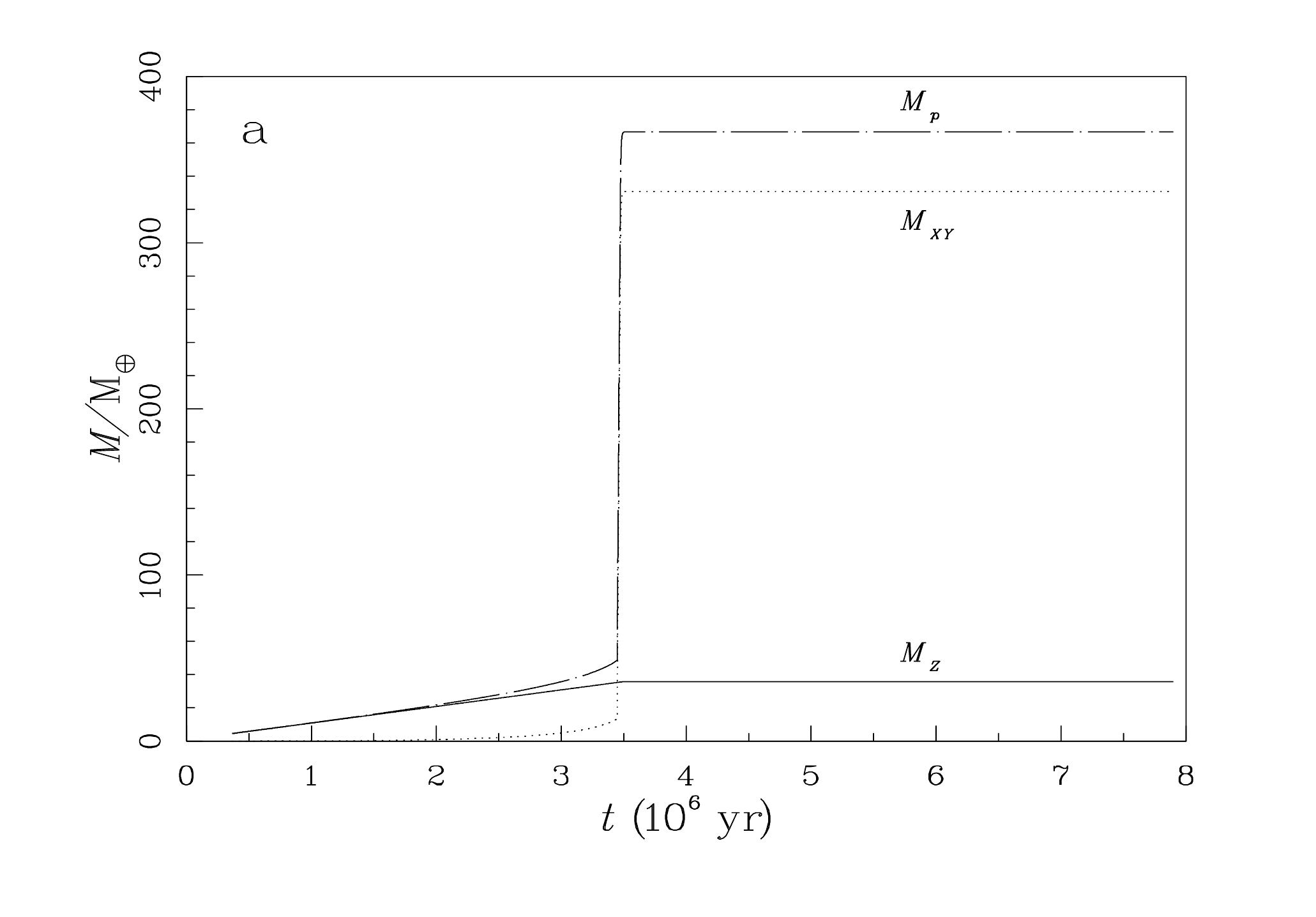}%
\includegraphics[trim={1cm 1cm 6cm 6cm},clip]{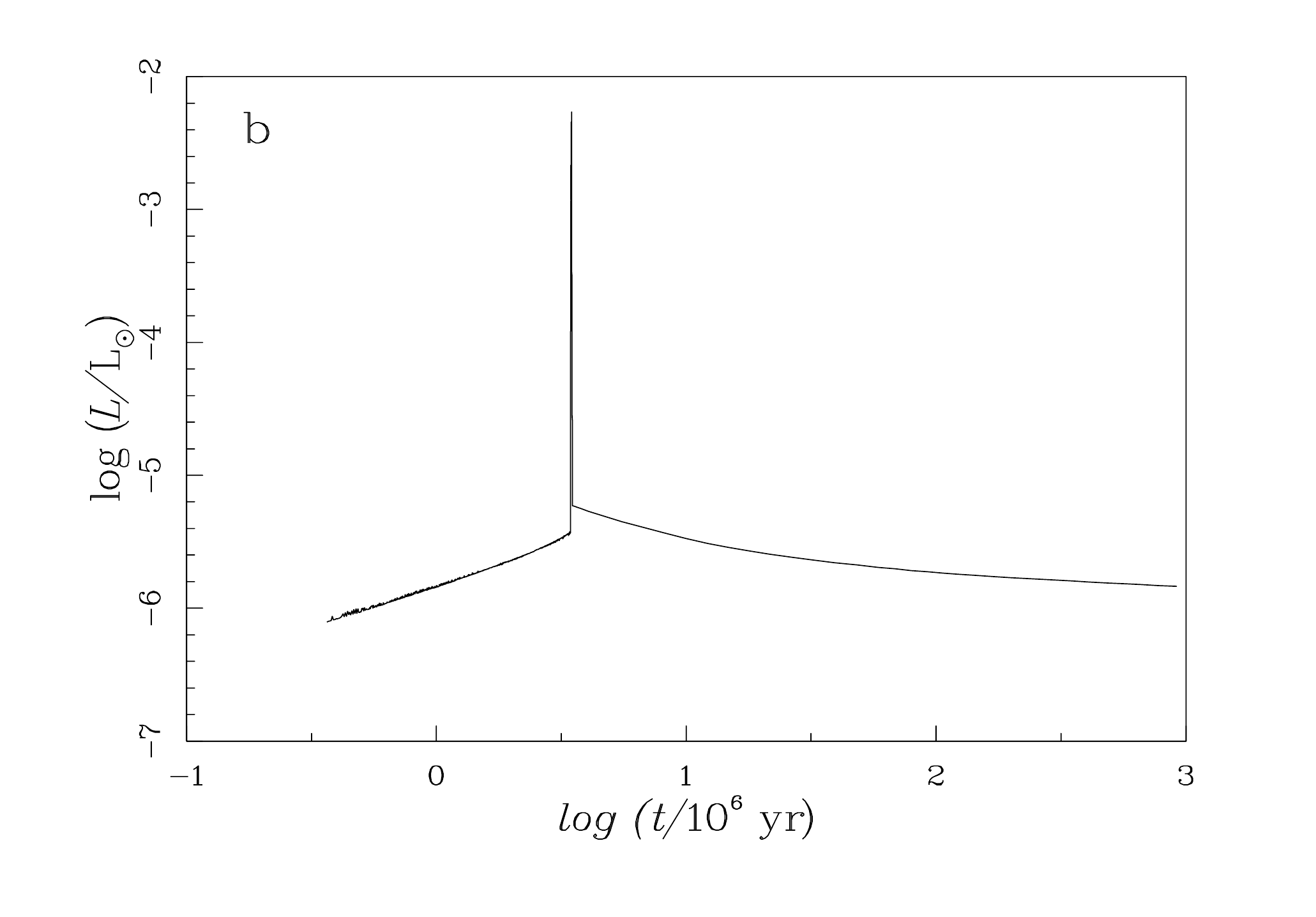}}
\caption{
The evolution of the companion to $\rho$ CrB in a 
standard protoplanetary disk and with constant solid accretion rate 
(Case C1). a) Mass of the solid component ({\it solid line}), 
the gaseous component ({\it dotted line}), and the total mass 
({\it dot-dashed line}) as  functions of time. 
b) Luminosity radiated by the planet as a function of time.
}
%\label{}
\end{figure*}

The core mass and envelope mass are shown as  functions of time in
Fig. 6a. The formation time to final mass is $3.59 \times 10^6$ yr, 
$M_{XY} = 331$ M\earth,  $M_{Z} = 36$ M\earth, and the total mass 
is 1.15 M$_J$.  The luminosity
as a function of time is shown in Fig. 6b. The peak value is almost
10$^{-2}$ L$_\odot$, higher than that in the inner planet case because
of the greater difference between the planetary 
radius and the accretion radius during the transition phase (Eq. 4) 
and the higher planetary mass. 
The width of the spike is $5 \times 10^4$ yr. Interior characteristics are
similar to those in the 51 Peg b model P1, with a maximum internal
temperature of 63,000 K. As a result of insolation in the isolated
phase, the luminosity levels off to log $L$/L$_\odot = -5.8$ and the surface 
temperature to 580 K. 

\section{CONCLUSIONS}

The  main purpose of this paper has been to demonstrate that
it is possible to model  the formation of 
the companions to 51 Peg, $\rho$ Cr B, and 
47 UMa  {\it in situ}, given  reasonable disk models, but that
there are problems associated with this assumption. 
Formation times of a few million 
years are obtained. For the companions  of  51 Peg and  $\rho$ Cr B the mass
accretion rate of solid material, $\dot M_Z$, is assumed to be constant. 
For the companion of 
47 UMa, cases with both constant and variable accretion rates are
considered. In the cases of constant $\dot M_Z$, the formation times are
comparable with observed disk lifetimes under the assumption that 
$\dot M_Z \approx  10^{-5}$ M\earth~ yr$^{-1}$. If that rate
were assumed to be an order of magnitude smaller, the formation times
would be $> 10^7$ yr (Bodenheimer and Pollack 1986), greater than
typical disk lifetimes. 
In the case of variable $\dot M_Z$ the crucial parameter is the disk surface
density of solid material, $\sigma_{init,Z}$. The location 
of 47 UMa b
at 2.1 AU corresponds to  probable nebula  temperatures above the
ice evaporation point; thus the available concentration of condensible
material is low. To obtain a reasonable formation time for the planet,
$\sigma_{init,Z}$ must be in the range 50 -- 100 g cm$^{-2}$, 
corresponding in standard disk models to $T \approx 1000$ K, and 
$\dot M_{neb} = 10^{-5}$ M$_\odot$ yr$^{-1}$. 
A change in the assumed value of $\sigma_{init,Z}$ from 50 to 90
g cm$^{-2}$ results in almost a factor 10 reduction in formation time, from
$1.86 \times 10^7$ to $1.94 \times 10^6$ yr. 
The instantaneous lifetime of such a nebula is much
shorter than the deduced formation time of the planet; however, the parameters
of the nebula are uncertain because the viscosity (assumed to 
correspond to $\alpha = 10^{-2}$) is not well constrained and, furthermore, 
the evolution time of the disk will soon slow down, with probable 
decoupling of the 
density of solids from that of the gas.
The initial tidal truncation
mass (Eq. 9) is 10 M$_J$ in this hot disk, but as the 
nebula evolves this limit will decrease, and, furthermore, the dispersal
of the nebula is another way to limit the planet's final mass.  
The conclusion can be reached that there are still difficulties in 
explaining the {\it in situ} formation
of 47 UMa b  by the models U1 and U2.
The main problem with the model is the length of
phase 2, during which both solid and gas accretion rates are low, 
unless $\sigma_{init,Z}$ is much larger than typical solar nebula
estimates would indicate. Such massive disks may be gravitationally unstable
(Boss 1998), indicating possible prior formation of a massive planet at 
larger distances. Similar problems would have been severe for the cases
of 51 Peg and $\rho$ CrB, had they been calculated with a variable 
accretion rate; however, the constant accretion rate that was used is
consistent with the estimates of Ward (1997b).  
The evolution of the solid and gaseous components of the disk
must be considered along with the planetary formation to make
further progress on these questions. 

Thus,   orbital
migration may well  be important. For  
51 Peg b, the case in which the gas density in
the inner part of the disk was highly reduced compared with standard
values is more likely to result in the survival of the planet than
the case which was calculated in a standard disk. Short-period 
planets which form in
regions of low gas density are likely to have  an enhanced solid component
for a given total mass; for
example, the core mass for the model of 51 Peg b in Case P1  is 47
M$_\oplus$, while in Case P3  it is 89 M$_\oplus$ out of a total mass, 
in both cases, of 167 M$_\oplus$.   
In the case of 47 UMa another possible scenario would involve the formation of
the planet at a larger distance, say 5 AU, where the models of 
Pollack {\it et al.} (1996) show that it is possible to form 
Jupiter-mass objects in 5~--~10 $\times~ 10^6$ yr in a disk with 
surface density about 3 times that of the minimum-mass solar nebula. 
Then migration could take it in to 2.1 AU. In fact one must consider
migration as a critical part of the formation process itself, since
migration begins long before the planet is massive enough
to open up a gap (Goldreich and Tremaine 1980; Ward 1997a,b). When
the mass of a protoplanet at Jupiter's orbit is 3 M$_\oplus$,  the
migration time is already as short as 10$^5$ yr (Ward 1997a,b). As the
mass approaches the nominal `isolation' mass, the migration of
the planet has carried it into regions of the disk where planetesimals
are still plentiful (Ward 1997b), or where there are other embryos with which
it can merge. Thus, the lengthy semi-isolated phase 2 in the Case U1
scenario could be avoided and the formation process could be speeded 
up considerably. An examination of these effects must
be included in future calculations. 
The problem still remains, as with all scenarios
for giant planet formation, of how to stop the migration at the observed
locations, especially for those planets farther than a tenth of an AU  
from the star that they orbit.

\appendix
\section{GLOSSARY OF SYMBOLS}
 
  $a \equiv$ the distance to the central star

  $c \equiv$ sound speed in the disk    

  $F_g \equiv$ gravitational enhancement factor for accretion of planetesimals  

  $g \equiv$ acceleration of gravity at $R_p$  

  $L \equiv$ radiated luminosity at surface of protoplanet  

  $L_{acc} \equiv$ luminosity due to gas accretion

  $L_{contr} \equiv$ luminosity due to contraction and accretion of solids

  $L_r \equiv$ energy per second crossing sphere of radius $r$  

  L$_\odot \equiv$ solar luminosity = $3.85 \times 10^{33}$ ergs/s  

  M$_\oplus \equiv$ Earth mass = $6 \times 10^{27}$ g  

  M$_{J} \equiv$ Jupiter mass = $1.899 \times 10^{30}$ g  

  M$_\odot \equiv$ solar mass = $1.989 \times 10^{33}$ g  

  $M_{cross} \equiv$ planet's $M_{Z}$ at which $M_{XY} = M_{Z}$  

  $M_p \equiv$ total mass of protoplanet  

  $M_{p,max} \equiv$ maximum mass that a planet can accrete before gap formation

  $M_\star \equiv$ mass of central star  

  $M_{XY} \equiv$ mass of gaseous envelope of protoplanet  

  $M_Z \equiv$ mass of solid core of protoplanet  

  $\dot M_{neb} \equiv$ accretion rate of disk mass onto central star  

  $\dot M_{XY}\equiv$ accretion rate of gas onto envelope  

  $\dot M_Z \equiv$ accretion rate of planetesimals onto protoplanet 

  $\dot M_{XY,max} \equiv$ limiting accretion rate of gaseous material

  $P \equiv$ photospheric pressure  

  $r \equiv$ distance from center of protoplanet

  $R_a \equiv$ modified accretion radius, Eq. (3)  

  $R_c \equiv$ effective capture radius of the protoplanet  

  $R_{core} \equiv$ radius at core/envelope interface

  $R_H \equiv$ protoplanet's tidal radius  

  $R_p \equiv$ actual outer radius of protoplanet  

  R$_\odot ~\equiv$ solar radius = $6.96 \times 10^{10}$ cm  

  $t \equiv$ time

  $T_c \equiv$ temperature of the gas within protoplanet at core/envelope interface  

  $T_{eff} \equiv$ $(L/4\pi R^2_p \sigma_{SB})^{1/4}$  

  $T_{neb} \equiv$ assumed temperature at $R_a$         

  $T_s \equiv$ temperature at $R_p$

  $v_{ff} \equiv$ free-fall velocity from $R_a$ to $R_p$  

  $X \equiv$ mass fraction of hydrogen  

  $Y \equiv$ mass fraction of helium  

  $Z \equiv$ mass fraction of elements heavier than helium  

  $\alpha \equiv$ viscosity parameter in disk model  

  $\kappa \equiv$ Rosseland mean opacity in units of cm$^2$/g  

  $\nu \equiv$ disk viscosity  

  $\rho_c \equiv$ density of the gas within the protoplanet at core/envelope interface  

  $\rho_{core} \equiv$ assumed density for core of protoplanet

  $\rho_{neb} \equiv$ assumed density at $R_a$          

  $\rho_s \equiv$ density at $R_p$  

  $\rho_0 \equiv$ density at the inner edge of the infalling flow  

  $\sigma_{init,XY} \equiv$ initial surface density of gaseous material in the disk

  $\sigma_{init,Z} \equiv$ initial surface density of condensed material in the disk

  $\sigma_{SB} \equiv$ Stefan-Boltzmann constant

  $\sigma_Z \equiv$ surface density of condensed material in the disk 

  $\Omega \equiv$ orbital frequency of protoplanet  

\section*{ACKNOWLEDGEMENTS}

This work was supported in part through NASA Grant NAG5-4494 from 
the Origins of Solar Systems Program. 
We dedicate this paper to the memory of Jim Pollack, 
in acknowledgment of his vision, wisdom, and hard work in connection 
with the problem of the origin of giant planets.

\section*{REFERENCES}
\medskip
\rfnce{
 Adams, F. C., and W. Benz 1992. Gravitational instabilities in 
circumstellar disks and the formation of binary companions. 
In {\it Complementary Approaches to
Double and Multiple Star Research} (H. McAlister and  W. Hartkopf, Eds.),
    pp. 185-194. Astronomical Society of the Pacific, San Francisco. 
}\rfnce{
 Alexander, D. R., and J. W. Ferguson 1994. Low-temperature 
Rosseland opacities. {\it Astrophys. J.} {\bf 437}, 879--891.
}\rfnce{
 Allard, F., P. H. Hauschildt, I. Baraffe, and G. Chabrier 1996. 
Synthetic spectra and mass determination of the brown dwarf Gliese 229B. 
{\it Astrophys. J.} {\bf 465}, L123--L127.
}\rfnce{
Artymowicz, P. 1992. Dynamics of binary and planetary-system 
interaction with disks: eccentricity changes.  
{\it Publ. Astron. Soc. Pacif.} {\bf 104}, 769--774. 
}\rfnce{
Artymowicz, P. 1998. On the formation of eccentric superplanets. 
In {\it Brown Dwarfs and Extrasolar Planets, ASP Conference Series Vol. 134}  
(R. Rebolo, E. L. Martin, and M. R. Zapatero Osorio, Eds.), pp. 152--161. 
Astronomical Society of the Pacific, San Francisco.
}\rfnce{
Artymowicz, P., and S. H. Lubow 1996. Mass flow through gaps in circumbinary 
disks. {\it Astrophys. J.} {\bf 467}, L77--L80. 
}\rfnce{
Beckwith, S. V. W., and A. I. Sargent 1993.     
The occurrence and properties of disks around  young stars. 
In {\it Protostars and 
Planets III} (E. Levy and J. Lunine,  Eds.), pp. 521-541.      
Univ. of Arizona Press, Tucson.                  
}\rfnce{
Bell, K. R., P. M. Cassen, H. H. Klahr, and Th. Henning 1997. The 
structure and appearance of protostellar accretion disks: limits
on disk flaring. {\it Astrophys. J.} {\bf 486}, 372--387. 
}\rfnce{
Black, D. C. 1997. Possible observational criteria for distinguishing 
brown dwarfs from planets. {\it Astrophys. J.} {\bf 490}, L171--L174.
}\rfnce{
Bodenheimer, P. 1998. Formation of substellar objects orbiting stars. 
In {\it Brown Dwarfs and Extrasolar Planets, ASP Conference Series Vol. 134}  
(R. Rebolo, E. L. Martin, and M. R. Zapatero Osorio, Eds.), pp. 115--127. 
Astronomical Society of the Pacific, San Francisco.
}\rfnce{
Bodenheimer, P., and J. B. Pollack 1986. Calculations of the accretion
and evolution of giant planets: The effects of solid cores. 
{\it  Icarus} {\bf 67}, 391--408.
}\rfnce{
Boss, A. P. 1996.  Evolution of the solar nebula. III. Protoplanetary disks 
undergoing mass accretion. {\it Astrophys. J.} {\bf 469}, 906--920.   
}\rfnce{
Boss, A. P. 1997.  Giant planet formation  by 
gravitational instability. {\it Science} {\bf 276}, 1836--1839. 
}\rfnce{
Boss, A. P. 1998. Evolution of the solar nebula. IV. Giant gaseous 
protoplanet formation. {\it Astrophys. J. } {\bf 503}, 923--937.   
}\rfnce{
Burrows, A., M. Marley, W. B. Hubbard, J. I. Lunine, T. Guillot,
D. Saumon, R. Freedman, D. Sudarsky, and C. Sharp 1997. A nongray
theory of extrasolar giant planets and brown dwarfs. 
{\it Astrophys. J.} {\bf 491}, 856--875. 
}\rfnce{
Butler, R. P., and G. W. Marcy 1996. A planet orbiting 47 Ursae 
Majoris. {\it Astrophys. J.} {\bf 464}, L153--L156.
}\rfnce{
Butler, R. P., G. W. Marcy, E. Williams, H. Hauser,  and P. Shirts 1997. 
 Three new ``51 Pegasi-type" planets. 
{\it Astrophys. J.} {\bf 474}, L115-L118. 
}\rfnce{
Cameron, A. G. W. 1978. Physics of the primitive
solar accretion disk. {\it Moon and Planets} {\bf 18}, 5--40. 
}\rfnce{
Cochran, W. D., A. P. Hatzes, R. P. Butler, and G. W. Marcy 1997. 
The discovery of a planetary companion to 16 Cygni B. 
{\it Astrophys. J.} {\bf 483}, 457--463. 
}\rfnce{
DeCampli, W. M., and A. G. W.  Cameron 1979. Structure and evolution
of isolated giant gaseous protoplanets. {\it  Icarus} 
{\bf 38}, 367--391. 
}\rfnce{
Goldreich, P., and S. Tremaine 1980.  Disk--satellite interactions.      
{\it  Astrophys. J. } {\bf 241}, 425--441. 
}\rfnce{
Greenzweig, Y., and J. J. Lissauer 1992. Accretion rates of protoplanets. 
II. Gaussian distributions of planetesimal velocities. {\it  Icarus}
{\bf 100}, 440--463. 
}\rfnce{
Guillot, T., D. Gautier, and W. B. Hubbard 1997. New constraints on
the composition of Jupiter from Galileo measurements and interior 
models. {\it  Icarus} {\bf 130}, 534--539. 
}\rfnce{
Holman, M, J. Touma, and S. Tremaine 1997. Chaotic variations in   
the eccentricity of the planet orbiting 16 Cygni  B. 
{\it Nature} {\bf 386}, 254--256. 
}\rfnce{
Kary, D. M., and J. J. Lissauer 1994. Numerical simulations of planetary 
growth. In {\it Numerical Simulations in Astrophysics} (J. Franco, 
S. Lizano, L. Aguilar, and E. Daltabuit, Eds.), pp. 364--373. 
Cambridge Univ. Press, Cambridge. 
}\rfnce{
K\"onigl, A. 1991. Disk accretion onto magnetic T Tauri 
stars.  {\it Astrophys. J.} {\bf 370}, L39--L43.
}\rfnce{
Kuiper, G. P. 1951. On the origin of the solar system. In 
{\it Astrophysics} (J. A. Hynek, Ed.), pp. 357--424. McGraw Hill, New York. 
}\rfnce{
Latham, D. W., T. Mazeh, R. Stefanik, M. Mayor, and G. Burki 1989. 
The unseen companion to HD114762: a probable brown dwarf. 
{\it Nature} {\bf 339}, 38--40.
}\rfnce{
Laughlin, G., and P. Bodenheimer 1994. Nonaxisymmetric evolution in
protostellar disks. {\it Astrophys. J.} {\bf 436}, 335--354. 
}\rfnce{
Levison, H. F., J. J. Lissauer, and M. J. Duncan 1998. Modeling the 
diversity of outer planetary systems. {\it Astron. J.} {\bf 116}, 1998--2014. 
}\rfnce{
Lin, D. N. C. 1997. Planetary formation in protostellar disks. In 
{\it Accretion Phenomena and Related Outflows}, IAU Colloquium No. 163 
 (D. T. Wickramasinghe,
G. V. Bicknell, and L. Ferrario, Eds.), pp. 321--330. Astronomical 
Society of the Pacific, San Francisco. 
}\rfnce{
Lin, D. N. C., P. Bodenheimer, and D. Richardson 1996. Orbital migration
of the planetary companion of 51 Pegasi to its present location. 
{\it Nature } {\bf 380}, 606--607. 
}\rfnce{
Lin, D. N. C., and S. Ida 1997. On the origin of massive eccentric
planets. {\it Astrophys. J. } {\bf 477}, 781--791. 
}\rfnce{
Lin, D. N. C., and J. C. B. Papaloizou 1979. Tidal torques on accretion
disks in binary systems with extreme mass ratios. 
{\it Mon. Not. R. Astron. Soc.} {\bf 186}, 799--812. 
}\rfnce{
Lin, D. N. C., and J. C. B. Papaloizou 1985. 
On the dynamical origin of the solar system. In {\it Protostars and 
Planets  II} (D. C. Black and M. S. Matthews, Eds.), pp. 981--1072. 
Univ. of Arizona Press, Tucson.                  
}\rfnce{
Lin, D. N. C., and J. C. B. Papaloizou 1986a. On the tidal interaction  
between protoplanets and the primordial solar nebula. II. Self-consistent 
non-linear interaction.                             
{\it Astrophys. J. } {\bf 307}, 395--409. 
}\rfnce{
Lin, D. N. C., and J. C. B. Papaloizou 1986b.  On the tidal interaction
between protoplanets and the protoplanetary disk. III. Orbital 
migration of protoplanets.                              
{\it Astrophys. J. } {\bf 309}, 846--857. 
}\rfnce{
Lin, D. N. C., and J. C. B. Papaloizou 1993. On the tidal interaction  
between protostellar disks and companions. In {\it Protostars and 
Planets III} (E. H. Levy and J. I. Lunine, Eds.), pp. 749--835. 
Univ. of Arizona Press, Tucson. 
}\rfnce{
Lissauer, J. J. 1987. Timescales for planetary accretion and 
the structure of the protoplanetary disk. {\it Icarus}
{\bf 69}, 249--265.
}\rfnce{
Lissauer, J. J. 1993. Planet formation. {\it Annu. Rev. Astron. Astrophys.}
{\bf 31}, 129--174. 
}\rfnce{
 Low, C., and D. Lynden-Bell 1976. The minimum Jeans mass, or when
fragmentation must stop.  {\it Mon. Not. R. Astron. Soc.} 
 {\bf 176}, 367--390. 
}\rfnce{
Marcy, G. W., and R. P. Butler  1996. A planetary companion to 70 
Virginis. {\it Astrophys. J.} {\bf 464}, L147--L151.
}\rfnce{
Marcy, G. W., and R. P. Butler  1998.  Detection of extrasolar 
giant planets.         
{\it Annu. Rev. Astron. Astrophys.} {\bf 36}, 57--97.  
}\rfnce{
Marley, M. S., D. Saumon, T. Guillot, R. S. Freedman, W. B. Hubbard, 
A. Burrows, and J. I. Lunine 1996. Atmospheric, evolutionary, and 
spectral models of the brown dwarf Gliese 229B. {\it Science} {\bf 272}, 1919--1921.
}\rfnce{
Mayor, M., and D. Queloz 1995. A Jupiter-mass companion to a 
solar-type star.  {\it Nature} {\bf 378}, 355--359. 
}\rfnce{
Mazeh, T., Y. Krymolowski, and G. Rosenfeld 1997. The high eccentricity 
of the planet orbiting 16 Cygni  B. 
{\it Astrophys. J.} {\bf 477}, L103--L106. 
}\rfnce{
Mizuno, H. 1980. Formation of the giant planets. {\it Prog. Theor. 
Phys.} {\bf 64}, 544--557. 
}\rfnce{
Murray, N., B. Hansen, M. Holman, and  S. Tremaine 1998. Migrating planets. 
{\it Science} {\bf 279}, 69--72.
}\rfnce{
Nakajima, T., B. R. Oppenheimer,  S. R.  Kulkarni, 
D. A. Golimowski, K.  Matthews, and S. T. Durrance 1995. Discovery of
a cool brown dwarf. {\it Nature}
 {\bf  378}, 463--465. 
}\rfnce{
Noyes, R. W., S. Jha, S. G. Korzennik, M. Krockenberger, P. 
Nisenson, T. Brown, E. Kennelly, and S. Horner 1997. 
A planet orbiting the star $\rho$ Coronae Borealis. 
{\it Astrophys. J.} {\bf 483}, L111--L114.
}\rfnce{
Oppenheimer, B. R., S. R. Kulkarni, K.  Matthews, and T. Nakajima
 1995.  Infrared spectrum of the cool brown dwarf Gl 229B.
{\it Science} {\bf 270}, 1478--1479. 
}\rfnce{
Perri, F., and A. G. W. Cameron 1974. Hydrodynamic instability of 
the solar nebula in the presence of a planetary core. {\it Icarus}
 {\bf 22}, 416--425. 
}\rfnce{
Podolak, M., J. B. Pollack, and R. T. Reynolds 1988. Interactions of
planetesimals with protoplanetary atmospheres. {\it Icarus}
 {\bf 73}, 163--179. 
}\rfnce{
 Pollack, J. B., O. Hubickyj, P. Bodenheimer, J. J. Lissauer, M. 
Podolak, and Y. Greenzweig 1996. Formation of the giant planets by 
concurrent accretion of solids and gas. {\it Icarus} {\bf 124}, 62--85.  
}\rfnce{
Pollack, J. B., C. McKay, and B. Christofferson 1985. A calculation of
the Rosseland mean opacity of dust grains in primordial Solar System
nebulae. {\it Icarus} {\bf 64}, 471--492.  
}\rfnce{
Rasio, F. A., and E. B. Ford 1996. Dynamical instabilities and the 
formation of extrasolar planetary systems. {\it Science} {\bf 274}, 954--956. 
}\rfnce{
 Rees, M. J. 1976. Opacity-limited hierarchical fragmentation and 
the masses of protostars. {\it Mon. Not. R. Astron. Soc.} {\bf 176}, 483--486.
}\rfnce{
 Ruzmaikina, T. V. 1998. Formation of 51 Peg type systems.        
{\it Lunar and Planetary Science Abstracts} {\bf 29}, 1873.     
}\rfnce{
Safronov, V. S. 1969. {\it Evolution of the Protoplanetary Cloud and 
Formation of the Earth and Planets}. Nauka Press, Moscow (in Russian). 
English translation: NASA--TT--F--677, 1972. 
}\rfnce{
Saumon, D., G. Chabrier, and H. M. Van Horn 1995. An equation of state
for low-mass stars and giant planets. 
{\it Astrophys. J. Suppl. } {\bf 99}, 713--741. 
}\rfnce{
Saumon, D., W. B. Hubbard, A. Burrows, T. Guillot, J. I. Lunine, 
and G. Chabrier 1996.  A theory of extrasolar giant planets. 
{\it Astrophys. J. } {\bf 460}, 993--1018. 
}\rfnce{
Shu, F. H., J. Najita, E. Ostriker, F. Wilkin, S. Ruden, and S. Lizano 1994. 
Magnetocentrifugally driven flows from young stars and disks. I. A generalized
model. {\it Astrophys. J. } {\bf 429}, 781--796. 
}\rfnce{
Stevenson, D. J. 1982. Formation of the giant planets. {\it Planet. Space 
Sci.} {\bf 30}, 755--764. 
}\rfnce{
Stringfellow, G. S., D. C. Black, and P. Bodenheimer 1990. Brown dwarfs 
as close companions to white dwarfs. {\it Astrophys. J.} {\bf 349}, L59--L62. 
}\rfnce{
Strom, S. E., S. Edwards, and M. F. Skrutskie 1993.  In {\it Protostars 
and Planets III} ( E. H. Levy and J. I. Lunine, Eds.), pp. 837--866. Univ. 
of Arizona Press, Tucson. 
}\rfnce{
Tajima, N., and Y. Nakagawa 1997. Evolution and dynamical stability
of the proto-giant-planet envelope. {\it Icarus} {\bf 126}, 282--292. 
}\rfnce{
Takeuchi, T., S. Miyama, and D. N. C. Lin 1996. Gap formation in protoplanetary 
disks. {\it Astrophys. J.} {\bf 460}, 832--847. 
}\rfnce{
Trilling, D., W. Benz, T. Guillot, J. I. Lunine, W. B. Hubbard, and
A. Burrows 1998. Orbital evolution and migration of giant planets: 
modeling extrasolar planets. {\it Astrophys. J.}  {\bf 500}, 428--439.
}\rfnce{
 Ward, W. R. 1997a. Survival of planetary systems.                          
{\it Astrophys. J.} {\bf 482}, L211--L214.  
}\rfnce{
 Ward, W. R. 1997b. Protoplanet migration by nebular tides.                 
{\it Icarus} {\bf 126}, 261--281.   
}\rfnce{
 Weidenschilling. S. J., and F. Marzari 1996. Gravitational scattering 
as a possible origin for giant planets at small stellar distances.  
{\it Nature} {\bf 384}, 619--621.  
}\rfnce{
 Wuchterl, G. 1991. Hydrodynamics of giant planet formation. III.          
Jupiter's nucleated instability.  {\it Icarus} {\bf 91}, 53--64. 
}\rfnce{
 Wuchterl, G. 1993. The critical mass for protoplanets revisited:         
Massive envelopes through convection. {\it Icarus} {\bf 106}, 323--334.
}\rfnce{
 Wuchterl, G. 1995. Giant planet formation. 
{\it Earth, Moon, and Planets} {\bf 67}, 51--65. 
}\rfnce{
 Wuchterl, G. 1997. Giant planet formation and the masses of extrasolar
planets. In {\it Science with the VLT Interferometer} (F. Paresce, Ed.), 
pp. 64--71. Springer, Berlin. 
}\rfnce{
Wuchterl, G., T. Guillot, and J. J. Lissauer  1999.  In {\it Protostars 
and Planets IV} (V. Mannings, A. P. Boss, and S. Russell, Eds.), 
in press.  Univ. of Arizona Press, Tucson. 
}

\printcredits

%%++++++++++++++++++++++++++++++++++++++++++++++++++++++++++++++++++++++%%
%%++++++++++++++++++++++++++++++++++++++++++++++++++++++++++++++++++++++%%

\end{document}